\documentclass[twocolumn,trackchanges,twocolappendix]{aastex7}
\usepackage{CJK}
\usepackage{graphicx}
\usepackage{natbib}
\usepackage{gensymb}
\usepackage{supertabular}
\hypersetup{linkcolor=blue,citecolor=blue,filecolor=blue,urlcolor=blue}

\usepackage{longtable}
\usepackage{threeparttablex}
\usepackage{multirow}
\usepackage{footmisc}
\usepackage{rotating}

\usepackage{booktabs}
\usepackage{caption, subcaption}

\usepackage{amsmath}
\usepackage{verbatim}

\usepackage{relsize} 

\newcommand{\HI}{H{\sc~i}}

\newcommand{\PII}{P{\sc~ii}}
\newcommand{\OI}{O{\sc~i}}
\newcommand{\AlII}{Al{\sc~ii}}
\newcommand{\SiII}{Si{\sc~ii}}
\newcommand{\SiIII}{Si{\sc~iii}}
\newcommand{\SiIV}{Si{\sc~iv}}
\newcommand{\SII}{S{\sc~ii}}
\newcommand{\CII}{C{\sc~ii}}

\newcommand{\CIV}{C{\sc~iv}}

\newcommand{\FeII}{Fe{\sc~ii}}
\newcommand{\NV}{N{{\sc~v}}}

\newcommand{\rvir}{R_{\rm 200m}}
\newcommand{\rvirm}{R_{\rm 200m}}
\newcommand{\rvirc}{R_{\rm 200c}}
\newcommand{\rproj}{R_{\rm \perp}}

\newcommand{\kms}{\rm km~s^{-1}}
\newcommand{\mhalo}{M_{\rm h}}

\newcommand{\mstar}{M_*}

\newcommand{\msun}{M_\odot}

\defcitealias{faerman_cool_2023}{FW23}

\begin{document}
\begin{CJK*}{UTF8}{gbsn} 

\title{The Halo Gas of Local Spiral Galaxies and the Link to Gaseous Satellites}

\author[orcid=0000-0003-4158-5116]{Yong Zheng (郑永)}
\affiliation{Department of Physics, Applied Physics and Astronomy, Rensselaer Polytechnic Institute, Troy, NY 12180, USA}
\email[show]{zhengy14@rpi.edu}

\author[orcid=0000-0002-9001-6713]{Jingyao Zhu (朱婧尧)}
\affiliation{Department of Astronomy, Columbia University, New York, NY 10027, USA}
\email{jingyao.zhu@columbia.edu}

\author[orcid=0009-0005-7307-5438]{Ben Wertz}
\affiliation{Department of Physics, Applied Physics and Astronomy, Rensselaer Polytechnic Institute, Troy, NY 12180, USA}
\email{wertzj2@rpi.edu}

\author[orcid=0000-0003-3520-6503]{Yakov Faerman}
\affiliation{School of Physics and Astronomy, Tel Aviv University, Tel Aviv 69978, Israel}\email{yakov.faerman@gmail.com}
\affiliation{Department of Astronomy, University of Washington, Seattle, WA 98195, USA}
\email{yakov.faerman@gmail.com}

\author[orcid=0000-0002-1129-1873]{Mary E. Putman}
\affiliation{Department of Astronomy, Columbia University, New York, NY 10027, USA}
\email{mputman@astro.columbia.edu}

\begin{abstract}

Gaseous satellites orbiting massive host galaxies experience gas stripping and may contribute additional gas to the hosts' circumgalactic medium (CGM). We identify a sample of 21 local spiral galaxies ($<$15 Mpc) and characterize their gaseous satellite populations from existing \HI\ surveys to investigate the connection between host galaxy CGM and satellite gas content. Most of our spiral hosts have $\leq3$ gaseous satellites in their halos with $M_{\rm HI}\gtrsim10^{7}~\msun$. Using 26 HST/COS QSO sightlines at impact parameters of 0.1--0.9 $R_{\rm 200c}$, we find that the CGM of $z\sim0$ spiral galaxies show large intrinsic scatters ($\sim1-2$ dex) in ion column densities and harbor a total cool gas mass of $2.5^{+7.5}_{\rm -1.9}\times10^9~(0.3Z_\odot/Z')~\msun$, largely consistent with their $z\sim0.2$ counterparts (e.g., COS-Halos). Splitting our sample by the presence of gaseous satellites, we find that galaxy hosts with gaseous satellites have higher detection rates (up to 50\%) in metal ion absorbers, including \OI, \AlII, \CII, \SiII, \SiIII, and \SiIV, possibly because their CGM is more metal enriched or has more ionized gas in the cool phase. However, the CGM column densities show no significant correlation with either the number of gaseous satellites or their total \HI\ masses, suggesting that the contribution from the gaseous satellites to the host CGM is likely small compared to the intrinsic scatters in the CGM profiles, and the profile trends are mainly influenced by sightline proximity to the host. The CGM detection rate becomes elevated when a sightline is within half the virial radius of a massive gaseous satellite (LMC-like or higher mass), likely due to stripped debris.

\end{abstract}



\section{Introduction} 

The circumgalactic medium (CGM) of a galaxy is baryon and metal rich and extends out to at least its virial radius \citep{ putman_gaseous_2012-2,tumlinson_circumgalactic_2017-1,chen26_cgm_review}. Satellite galaxies move through this halo medium as they orbit the galaxy, and can be quenched as their gas is stripped by the host's CGM \citep{zhu24,mayer_simultaneous_2006}.  The relationship between the CGM content of galaxies and the properties of their satellites can be best studied in the local Universe, where faint satellites are detectable and the proximity of the host halos makes it easier to locate background sightlines that probe the CGM.


Estimates of the CGM mass depend on the measured spatial profiles of multiphase ion absorbers and their covering fractions \citep[e.g.,][]{werk_cos-halos_2014, keeney17}. There have been many observational studies of the density of the CGM in various ions with radius from the center of a galaxy \citep[e.g.,][]{werk_cos-halos_2013, keeney17}. A typical L$_*$ galaxy has been found to host large quantities of baryons in its extended halo \citep{tumlinson_circumgalactic_2017-1, chen26_cgm_review}. However, the CGM of the Milky Way (MW) is a puzzle with its low amount of baryons in comparison to other L$_*$ galaxies \citep{zheng_characterizing_2020, bish_quastar_2021}. M31 also has a distinct CGM compared to some of the larger extragalactic samples such as the COS-Halos survey in that it lacks strong \HI\ absorbers in its inner halo, and its inferred cool CGM mass is lower \citep{howk17, lehner_project_2020, lehner26_amiga2}. 

Several factors could contribute to the differences in CGM among the MW, M31, and extragalactic samples.  For example, for the MW, the inside-out viewing angle of the CGM is a complicating factor \citep{zheng15, zheng_tentative_2019, bish_quastar_2021}.  For the case of M31, the ionized envelope of the Magellanic Stream \citep{fox14} covers nearly half of M31's CGM in the foreground \citep{lehner_project_2020, lehner26_amiga2}.   It is also possible there is a pervasive Local Group medium that affects the CGM measurements of the MW and M31 \citep{nuza14, putman_gas_2021}. And, $z\sim0$ galaxies may have different properties than their extragalactic counterparts at earlier times \citep[e.g.,][]{werk_cos-halos_2013, keeney17}.

Studies show that satellites at $z\sim0$ show a clear lack of gas in the vicinity of spiral galaxies and this is likely strongly linked to the host's CGM \citep{putman_gas_2021, zhu_census_2023, zhu_baryonic_2025}. Ram pressure stripping \citep{gunn_infall_1972-1}, the direct removal of a satellite's interstellar medium (ISM) via interaction with a host halo medium, is a primary quenching mechanism in simulated MW analogs (e.g., \citealt{simpson_quenching_2018,simons_figuring_2020,akins_quenching_2021,samuel_extinguishing_2022,engler_satellites_2023,rodriguez-cardoso_agora_2025}). 
Given this gas removal process, one might expect a correlation between the gaseous content of satellite galaxies and the host CGM that they are moving through. For instance, do hosts with less prominent halo content retain more gaseous satellites at $z \sim 0$ because stripping is less efficient? And in turn, how do stripped satellites contribute to the CGM properties of the hosts?

In this paper, we examine the CGM and gaseous satellite population of spiral galaxies at $z\sim0$. The paper is organized as follows. In Section \ref{sec:data_method}, we present our $z\sim0$ spiral galaxy sample and gaseous satellite census from existing \HI\ surveys, and describe our UV QSO sightline sample and relevant spectral analyses. In Section \ref{sec:result}, we present our results of observed CGM ion column density profiles of $z\sim0$ galaxies and compare them with low-$z$ samples from the literature. We also examine the CGM detections and non-detections in relation to the properties of the gaseous satellites of the hosts in this section.  Section \ref{sec:discuss_model} constructs an analytical CGM profile from our absorber results, and Section \ref{sec:discussion} calculates the total cool CGM mass for the $z\sim0$ spiral galaxies and discusses the presence of gaseous satellites in the context of the CGM results (\S\ref{sec:discuss_sat_pop}). We summarize the key takeaways in Section \ref{sec:summary}.  

\begin{figure}[t]
    \centering
    \includegraphics[width=\columnwidth]{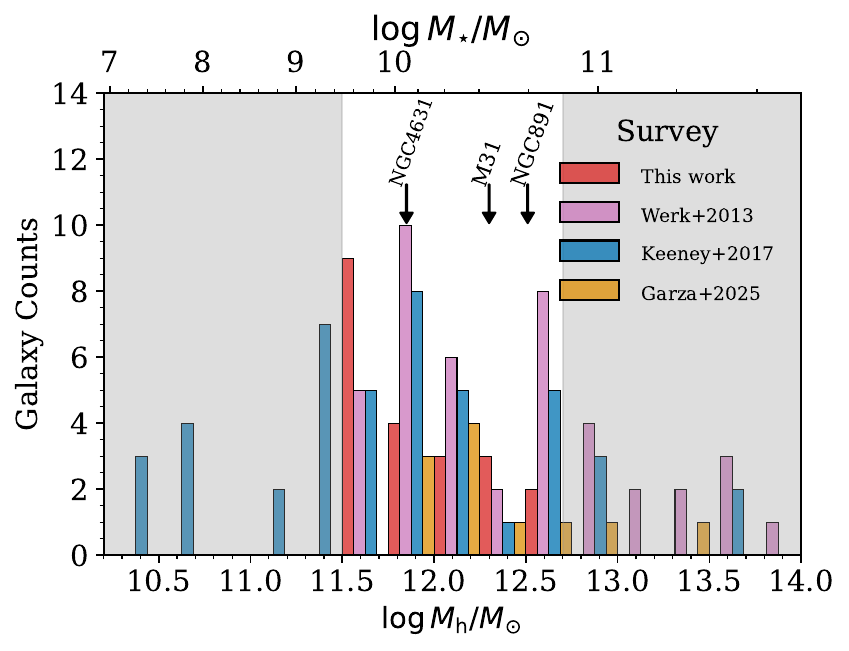}
    \caption{Histogram of the halo (bottom x-axis) and stellar (top) masses of our galaxy sample (red bars), compared to existing literature surveys. We focus on local ($z\sim0$) star-forming spirals in the halo mass range of $\mhalo=10^{11.5-12.7}~\msun$. See Section \ref{sec:gal_sample} for details. }
    \label{fig:gal_population}
\end{figure}

\section{Data And Methodology}
\label{sec:data_method}

\begin{deluxetable*}{cccccccccccc}
\tabletypesize{\footnotesize}
\tablecaption{Local Spiral Galaxy Sample \label{tb:gal_info}}  
\tablehead{
    \colhead{GID} &
    \colhead{Galaxy} &
    \colhead{$D$} &
    \colhead{$v_{\rm helio}$} &
    \colhead{$\log\mstar$} &
    \colhead{$\log\mhalo$} &
    \colhead{$\log M_{\rm HI}$} &
    \colhead{$\theta$} &
    \colhead{$\rvirc$} & 
    \colhead{\# of \HI} & 
    \colhead{Satellite} &
    \colhead{Dwarf $\log M_{\rm HI,lim}$}\\
& & [Mpc] & [$\kms$] & [$\log\msun$] & [$\log\msun$] & [$\log\msun$] & [$\degree$] & [kpc] &Satellites &  Source & [$\log\msun$]\\
\colhead{(1)} & \colhead{(2)} & \colhead{(3)} & \colhead{(4)} & \colhead{(5)}  & \colhead{(6)} & \colhead{(7)} & \colhead{(8)} & \colhead{(9)} & \colhead{(10)} & \colhead{(11)} & \colhead{(12)} 
 }
\startdata 
\hline \hline
1 & NGC3432$^{a}$ & 9.35 & 616 & 9.50 & 11.51 & 9.51 & 85 & 144.7 & 0 & F & $>$6.44\\
2 & NGC4517 & 8.30 & 1127 & 9.50 & 11.51 & 9.32 & 90 & 144.7 & 1 & A--H--E & 6.9\\
3 & NGC5102 & 3.83 & 467 & 9.58 & 11.55 & 8.48 & 90 & 149.5 & 0 & H & 6.83\\
4 & IC5332 & 8.79 & 706 & 9.65 & 11.59 & 9.50 & 30 & 153.9 & 0 & H & 7.55\\
5 & NGC5364 & 14.16 & 1268 & 9.67 & 11.60 & 9.39 & 52 & 155.2 & 2 & A & 7.23\\
6 & NGC1087 & 14.22 & 1523 & 9.83 & 11.69 & 9.23 & 53 & 166.0 & 0 & F--H & 7.31\\
7 & NGC3511 & 12.16 & 1109 & 9.87 & 11.71 & 9.48 & 73 & 169.0 & 1 & M--H & 5.7\\
8 & NGC4274 & 9.60 & 917 & 9.90 & 11.73 & 8.20 & 68 & 171.2 & 1 & A--F & 6.81\\
9 & NGC3486 & 9.70 & 677 & 9.90 & 11.73 & 9.41 & 46 & 171.2 & 7 & A & 6.88\\
10 & NGC4559 & 8.91 & 813 & 10.00 & 11.79 & 9.69 & 68 & 179.1 & 0 & A & 6.84\\
11 & NGC4026$^{b}$ & 13.67 & 985 & 10.26 & 11.95 & 0.00$^{b}$ & 90 & 203.7 & 0 & F & 6.77\\
12 & NGC5457 & 6.50 & 243 & 10.30 & 11.98 & 10.20 & 22 & 208.2 & 2 & F--E & 7.24\\
13 & NGC4736 & 4.20 & 287 & 10.30 & 11.98 & 8.65 & 38 & 208.2 & 0 & F--E & 6.26\\
14 & NGC3351$^{c}$ & 9.30 & 777 & 10.40 & 12.06 & 9.03 & 49 & 220.8 & 1 & A--D & 6.89\\
15 & NGC7814 & 13.55 & 1051 & 10.43 & 12.08 & 8.98 & 90 & 225.0 & 1 & A & 7.16\\
16 & NGC4535 & 14.83 & 1964 & 10.47 & 12.12 & 9.57 & 52 & 231.1 & 0 & A & 7.3\\
17 & NGC4258$^{d}$ & 7.20 & 462 & 10.60 & 12.25 & 9.80 & 73 & 254.7 & 6 & F--E--D & 7.35 \\
18 & NGC3368$^{c}$ & 9.90 & 893 & 10.60 & 12.25 & 9.26 & 49 & 254.7 & 1 & A--D & 6.87 \\
19 & NGC5194 & 8.60 & 465 & 10.70 & 12.37 & 9.74 & 33 & 279.4 & 1 & F--E & 7.08 \\
20 & NGC3627$^{e}$ & 9.40 & 716 & 10.80 & 12.51 & 8.92 & 66 & 313.0 & 7 & A--E & 6.77 \\
21 & NGC4565 & 11.90 & 1261 & 10.90 & 12.69 & 9.94 & 89 & 359.3 & 3 & A--E & 7.19 \\
\hline
\enddata
\tablenotetext{}{
Note: 
Col (1): Galaxy ID. 
Col (2): Galaxy name. 
Col (3): Distance. 
Col (4): Heliocentric velocity. 
Col (5): Stellar mass. 
Col (6): Halo mass based on the $z=0$ stellar mass-halo mass relation in \cite{behroozi_universemachine_2019}. 
Col (7): \HI\ mass. 
Col (8): Disk inclination angle.  
Col (9): Virial radius. 
Col (10): Number of gaseous satellites with confirmed optical counterparts within host galaxies' virial radii. 
Col (11): Data sources for the satellite census. \HI\ surveys: ``A" = ALFALFA \citep{haynes_arecibo_2011}, ``H" = HIPASS \citep{meyer_hipass_2004}, ``F" = FASHI \citep{zhang_fast_2024}, ``M": MHONGOOSE \citep{de_blok_mhongoose_2024}. Optical surveys: ``E" = ELVES \citep{carlsten_exploration_2022-1}, ``D" = Dragonfly \citep{cohen_dragonfly_2018}. See Section \ref{sec:gal_sample} for detail.
Col (12): Estimated \HI\ sensitivity to dwarf galaxies at the host location. See Section \ref{sec:gal_sample}.\\
$^{a}$ NGC 3432 is within the FASHI footprint but not included in \cite{zhang_fast_2024} due to local RFI issues; we recalculated its \HI\ properties using a cleaned FAST cube (C-P Zhang priv. comm.). Its estimated local sensitivity is likely a lower limit.\\
$^{b}$ NGC 4026 is a lenticular galaxy with no \HI\ emission. We found no gaseous satellite in FASHI, but previous work identified \HI\ in a tidally-disturbed companion \citep{appleton_extended_1983}.\\
$^{c}$ NGC 3351 (M95) and NGC 3368 (M96) are in the M96 or Leo I group \citep{muller_leo-i_2018} and of comparable masses. Satellites are assigned based on proximity to each host \citep{zhu_census_2023}.\\
$^{d}$ NGC 4258 (M106) is the brightest galaxy in the CVn II group \citep{fouque_groups_1992} with many bright satellite candidates (e.g., \citealt{spencer_survey_2014}).\\
$^{e}$ NGC 3627 (M66) is the primary of Leo Triplet (with NGC 3628 and NGC 3623) or the M66 group. The halo region and satellite content are more contaminated due to this group environment.\\
} 
\end{deluxetable*}

\subsection{Spiral Galaxy Sample and Gaseous Satellite Census} 
\label{sec:gal_sample}

Our sample includes 21 spiral galaxies in the local Universe ($<$15 Mpc) with stellar masses of $\mstar=10^{9.5-10.9}~\msun$ and halo masses of $\mhalo=10^{11.5 - 12.7}~\msun$ (see Table \ref{tb:gal_info}). The halo masses are derived from the median stellar mass-halo mass relation from the UniverseMachine DR1 \citep{behroozi_universemachine_2019}, adopting the parameterization for star-forming central galaxies at $z=0$. 
Figure \ref{fig:gal_population} shows the halo mass histogram of our galaxy sample compared to existing low-$z$ galaxy surveys including the COS-Halos \citep{werk_cos-halos_2013}, \cite{keeney17}, and the CIViL$^*$ survey \citep{garza25_civil}; we describe the details of these literature surveys in Section \ref{sec:uv_literature}. Our sample traces a uniform time in the Universe ($z\sim0$) for a specific type of galaxy (spiral), and thus reduces systemic uncertainties that are present in some of the literature studies. 

The galaxies are chosen to be approximate analogs to the MW and M31 in mass and morphology, and to have archival COS spectra in G130M and/or G160M gratings for QSO sightlines within their virial radii. We define the virial radius\footnote{Sometimes $\rvirm$ is used in the literature for the virial radius, where the average enclosed density is 200 times the mean matter density ($\Omega_{m}\rho_{\rm crit}$). $\rvirm$ is a constant factor of $\Omega_{m}^{-1/3} \approx 1.55$ larger than $\rvirc$ at $z \sim 0$.}, $\rvirc$, as the boundary within which the mean density is 200 times the critical density of the Universe at $z=0$ ($\mhalo = \frac{4 \pi}{3} \cdot 200 \rho_{\rm crit} \rvirc^{3}$). Throughout this work, we adopt cosmological parameters from \cite{planck18} through the \texttt{Astropy} package \citep{astropy:2022}.

Our focus on nearby galaxies enables not only more QSO sightlines because of the larger angular sizes of the halos, but also a detailed census of their dwarf satellite populations. In this work, we focus on the gas-bearing satellites to examine the relation between satellite gas content and the host CGM. We identify gas-bearing satellites around all 21 galaxies in our sample using wide-field \HI\ surveys --- FASHI \citep{zhang_fast_2024} and ALFALFA \citep{haynes_arecibo_2011} -- in the northern hemisphere, and HIPASS \citep{meyer_hipass_2004,doyle_hipass_2005} in the southern hemisphere. For one galaxy, NGC 3511, we use the targeted deep \HI\ observations from the MHONGOOSE survey \citep{de_blok_mhongoose_2024}. 
The sources of the satellite data are noted in Table \ref{tb:gal_info}.  For 9 of the 21 hosts, there is a census of the quenched (gas-poor) satellites from targeted optical studies that we also note \citep{cohen_dragonfly_2018,carlsten_exploration_2022-1}. 

\begin{figure}[t]
    \centering
    \includegraphics[trim={0.5cm 0cm 1.8cm 1.5cm}, clip, width=\columnwidth]{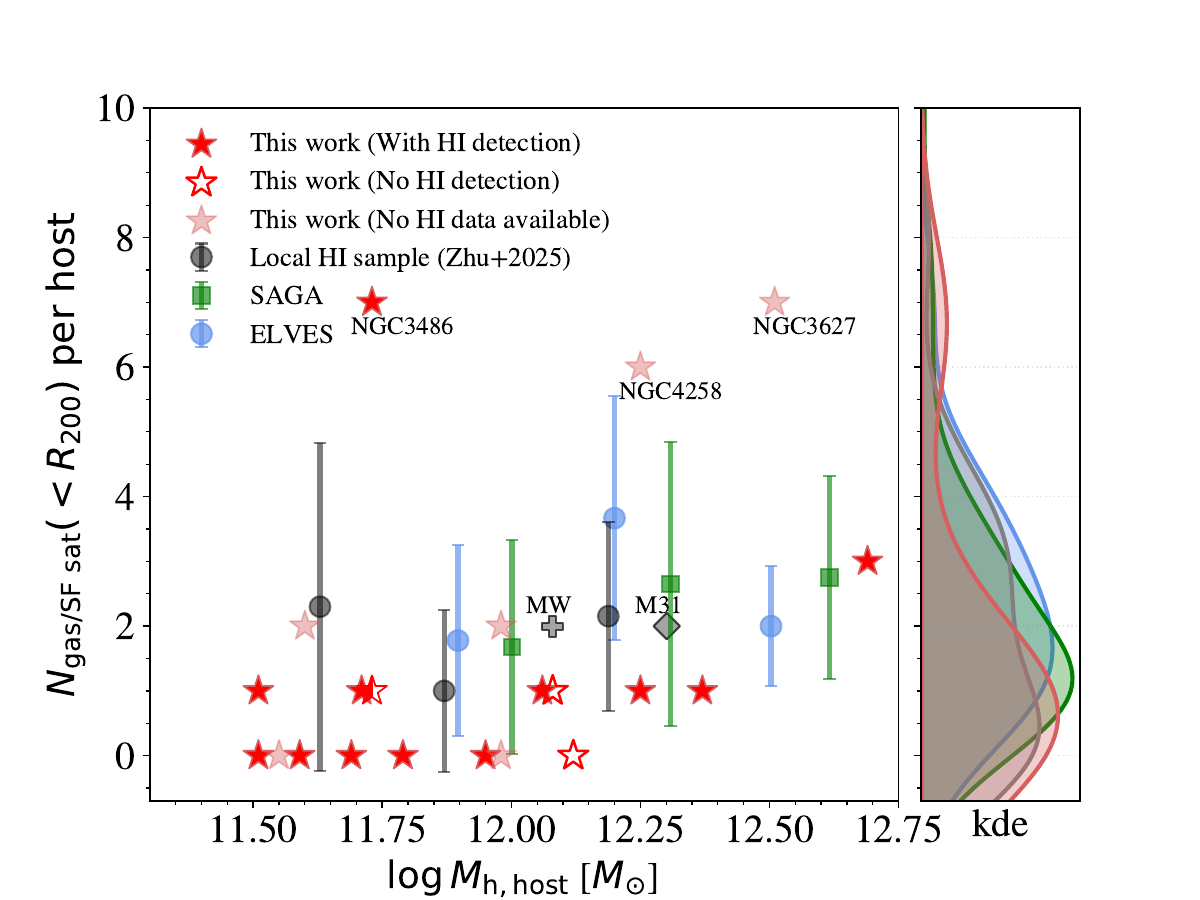}
    \caption{Number of \HI-bearing satellite galaxies within the virial radii ($\rvirc$) of our spiral host galaxies at $z \sim 0$. Host galaxies in our sample (Table \ref{tb:gal_info}; Section \ref{sec:gal_sample}) are shown as stars, with filled red stars indicating \HI\ Ly$\alpha$ detections in host galaxies' CGM, open red stars indicating non-detections, and pale red stars indicating that the sightlines do not have \HI\ measurements, either due to a lack of coverage (G160M only), or that there is a severe blending with the MW Ly$\alpha$ signals. One galaxy, NGC4274, is shown as a half-filled star near $x\sim11.75$ and $y=1$, which indicates a mixture of detection and non-detection from two sightlines (see Section \ref{sec:our_uv_sample}). Other color-coded symbols and errorbars show the binned averages and $\pm 1\sigma$ scatter from previous satellite studies, including a Local Volume \HI\ sample with $\log (M_{\rm h}/M_{\odot}) \geq 11.5$ (\citealt{zhu_baryonic_2025}; gray), the SAGA Survey  (\citealt{mao_saga_2024}; green), late-type hosts in the ELVES survey (\citealt{carlsten_exploration_2022-1}; blue), and the MW and M31 (\citealt{putman_gas_2021}; gray cross and diamond). The right panel shows the kernel density estimation (KDE) of the satellite counts.
    }
    \label{fig:sat_population}
\end{figure}

We identify gas-bearing satellite candidates from the wide-field \HI\ surveys using cuts in projected distance ($d_{\rm proj} \leq R_{200\rm c,host}$) and the line-of sight velocity differences between satellites and host galaxies ($\Delta V_{\odot \rm (sat-host)}(r) \leq V_{\rm esc,host}(r)$; see \citealt{zhu_baryonic_2025} for details), where $V_{\rm esc,host}$ is the point-mass escape velocity for the host, and $\Delta V_{\odot,\rm (sat-host)}$ is the heliocentric velocity difference between the two systems. We crossmatch these candidates with DESI Legacy Surveys DR10 \citep{dey_overview_2019} and only retain galaxies with optical counterparts to avoid \HI\ clouds, tidal features, and fragmented components of larger galaxies. To reduce contamination from interlopers, we use the cataloged redshift-independent distances of dwarf galaxies around the nine hosts with targeted optical data \citep{cohen_dragonfly_2018,carlsten_exploration_2022-1} to calibrate a more restrictive velocity selection criterion,
\begin{equation}
    \sqrt{3} \cdot |\Delta V_{\odot,\rm (sat-host)}| \leq V_{\rm esc,host}(\sqrt{3/2} \cdot d_{\rm proj}),
    \label{eq:vesc}
\end{equation} 
where $V_{\rm esc,host}$ is evaluated at the de-projected satellite-host separation $r=\sqrt{3/2} \cdot d_{\rm proj}$. This criterion provides a good match with the ELVES satellite/background classification based on surface brightness fluctuation distances.

In total, we find 34 gas-bearing satellite galaxies across the 21 spiral hosts, with a mean $5 \sigma$ \HI\ mass sensitivity of $M_{\rm HI,lim} \approx 10^{7}\ M_{\odot}$. The \HI\ mass sensitivity of each host is listed in Table \ref{tb:gal_info}, which is computed using the host distance and the local root-mean-square noise of the corresponding \HI\ survey, assuming a velocity linewidth at 50\% peak flux of $W_{50} = 25\ \kms$ (see Equation 3 in \citealt{zhu_baryonic_2025}). We also include gas-bearing companion galaxies more massive than dwarfs ($M_{\star} \geq 10^{9}\ M_{\odot}$), such as lower-mass spirals in galaxy groups where the host is the group central. Although these massive companions are often excluded from dwarf satellite studies \citep[e.g.,][]{zhu_census_2023}, we include them here to provide a complete census of the relatively massive gas-bearing satellite population. The full satellite catalog is provided in Appendix \ref{app3:gaseous_sats}.

We summarize the number of \HI-bearing satellites within $\rvirc$ versus the host halo mass in Figure \ref{fig:sat_population}. For reference, we include \HI-bearing satellites around the MW and M31 above our \HI\ mass sensitivity limit \citep{putman_gas_2021} and \HI-bearing or star-forming\footnote{For relatively massive dwarf galaxies, \HI-bearing and star-forming populations are largely equivalent \citep{meurer_survey_2006, lee_galex_2011}.} satellites around spiral galaxies at $z \sim 0$ from a Local Volume \HI\ survey \citep{zhu_baryonic_2025}, the SAGA survey \citep{mao_saga_2024}, and late-type hosts in the ELVES survey \citep{carlsten_exploration_2022-1}. All extragalactic samples are able to detect $\gtrsim 50\%$ of the equivalent Local Group dwarf galaxies. 

\begin{deluxetable*}{ccccccccc}
\tabletypesize{\footnotesize}
\tablecaption{Archival QSO Sight Lines Information \label{tb:qso_info}}
\tablehead{
    \colhead{ID} &
    \colhead{QSO Name} &
    \colhead{RA (J2000)} &
    \colhead{DEC (J2000)} &
    \colhead{Host Galaxy} &
    \colhead{$\rproj$} &
    \colhead{$\rproj/\rvirc$} &
    \colhead{UV Data} & 
    \colhead{SNR} \\
& & [$\degree$] & [$\degree$] & & [kpc] & & & \\
\colhead{(1)} & \colhead{(2)} & \colhead{(3)} & \colhead{(4)} & \colhead{(5)}  & \colhead{(6)} & \colhead{(7)} & \colhead{(8)} & \colhead{(9)} 
 }
\startdata 
\hline \hline
1 & CSO295 & 163.0232 & 36.6777 & NGC3432 & 16.9 & 0.12 & G130M & 9.2\\
2 & SDSSJ123304.05-003134.1 & 188.2669 & -0.5262 & NGC4517 & 93.3 & 0.64 & G130M+G160M & 9.1, 7.7\\
3 & IRAS13224-3809 & 201.3303 & -38.4149 & NGC5102 & 127.3 & 0.85 & G130M+G160M & 11.9, 5.7\\
4 & RBS2023 & 353.7187 & -35.6450 & IC5332 & 71.5 & 0.46 & G130M & 11.6\\
5 & HE2332-3556 & 353.6851 & -35.6631 & IC5332 & 68.1 & 0.44 & G130M & 14.9\\
6 & SDSS-J135726.27+043541.4 & 209.3595 & 4.5948 & NGC5364 & 128.7 & 0.83 & G130M+G160M & 9.8, 11.2\\
7 & J0246-0059 & 41.7163 & -0.9920 & NGC1087 & 125.5 & 0.76 & G130M+G160M & 7.1, 10.9\\
8 & PMNJ1103-2329 & 165.9067 & -23.4917 & NGC3511 & 86.7 & 0.51 & G130M+G160M & 13.8, 10.3\\
9 & SDSSJ110307.57+291230.0 & 165.7816 & 29.2085 & NGC3486 & 108.4 & 0.63 & G130M & 8.6\\
10 & SDSSJ110052.33+283801.2 & 165.2180 & 28.6337 & NGC3486 & 60.1 & 0.35 & G130M & 8.8\\
11 & QSO-B1215+303 & 184.4670 & 30.1168 & NGC4274 & 110.3 & 0.64 & G130M+G160M & 20.5, 21.2\\
12 & PG1218+304 & 185.3414 & 30.1770 & NGC4274 & 108.9 & 0.64 & G130M & 12.1\\
13 & SDSSJ123657.20+265701.2 & 189.2383 & 26.9503 & NGC4559 & 160.3 & 0.90 & G130M & 6.7\\
14 & 2MASS-J11590172+5106308 & 179.7572 & 51.1086 & NGC4026 & 38.0 & 0.19 & G130M+G160M & 2.9, 7.2\\
15 & SDSSJ140732.25+550725.6 & 211.8844 & 55.1237 & NGC5457 & 112.9 & 0.54 & G130M & 11.3\\
16 & 2MASS-J12493903+4122440 & 192.4128 & 41.3788 & NGC4736 & 25.5 & 0.12 & G160M & 10.5\\
17 & VV2006-J125901.7+413055 & 194.7570 & 41.5155 & NGC4736 & 115.7 & 0.56 & G130M+G160M & 9.9, 11.4\\
18 & J104335.86+115129.1 & 160.8995 & 11.8581 & NGC3351 & 28.9 & 0.13 & G130M & 11.6\\
19 & PG0003+158 & 1.4968 & 16.1636 & NGC7814 & 155.6 & 0.69 & G130M+G160M & 21.4, 24.1\\
20 & SDSSJ123426.80+072411.3 & 188.6117 & 7.4032 & NGC4535 & 205.8 & 0.89 & G130M & 6.6\\
21 & UVQSJ122208.10+461250.1 & 185.5338 & 46.2139 & NGC4258 & 153.0 & 0.60 & G160M & 6.5\\
22 & 2MASS-J12204664+4643478 & 185.1942 & 46.7298 & NGC4258 & 81.9 & 0.32 & G160M & 6.9\\
23 & J104709.83+130454.6 & 161.7910 & 13.0819 & NGC3368 & 218.4 & 0.86 & G130M & 7.6\\
24 & SDSSJ132222.68+464535.2 & 200.5945 & 46.7598 & NGC5194 & 202.8 & 0.73 & G130M+G160M & 6.2, 7.1\\
25 & 2MASS-J11230490+1257482 & 170.7705 & 12.9634 & NGC3627 & 113.2 & 0.36 & G160M & 6.2\\
26 & SDSSJ123604.02+264135.9 & 189.0165 & 26.6933 & NGC4565 & 147.2 & 0.41 & G130M+G160M & 9.1, 6.1\\
\hline
\enddata
\tablenotetext{}{
Note: 
Col (1): ID for each galaxy-QSO pair. 
Col (2): QSO name. 
Col (3): Right ascension. 
Col (4): Declination. 
Col (5): Host galaxy name. 
Col (6): Impact parameter (projected distance) of a QSO sight line from its host galaxy. 
Col (7): Impact parameter normalized to the host galaxy's virial radius. 
Col (8): HST/COS UV spectral gratings used in this work. 
Col (9): Spectral signal-to-noise ratio (SNR) per resolution element. When two numbers are listed, they indicate the SNRs for G130M and G160M gratings, respectively.  
} 
\end{deluxetable*}

Figure \ref{fig:sat_population} shows that local MW-like spiral galaxies typically host $\lesssim5$ \HI-bearing or star-forming satellites. The distribution of satellite counts peaks at low numbers ($\sim 1-2$; right panel) and is overall consistent across surveys, with most galaxies in our sample (18/21) hosting $\leq 3$ gaseous satellites. Differences among surveys are mild and mainly reflect variations in host halo mass (left panel) and satellite selection criteria. A small fraction of hosts in our sample (3/21; $\sim 14\%$) appear as ``outliers" with $>5$ gaseous satellites, slightly higher than previous studies ($\sim 10\%$; \citealt{carlsten_exploration_2022-1,mao_saga_2024,zhu_baryonic_2025}).  This is at least partially due to our inclusion of lower-mass spirals around hosts in galaxy groups. The outliers, identified in Figure \ref{fig:sat_population}, include two galaxy group centrals with more complex environments (NGC 4258 and NGC 3627; Table \ref{tb:gal_info}) and one previously identified outlier (NGC 3486; \citealt{zhu_baryonic_2025}). In contrast to the low number of gaseous or star-forming satellites, many hosts in our sample harbor a larger population of quiescent, gas-poor satellites (see \citealt{carlsten_exploration_2022-1} for details).

\subsection{Our Archival COS Sample and Analysis}
\label{sec:our_uv_sample}

Within the virial radii of our sample galaxies, we searched the MAST archive for publicly available HST/COS QSO sightlines as of January 22, 2025 and identified 26 archival QSO sightlines with sufficient signal-to-noise ratios (SNRs; see Table \ref{tb:qso_info}). These sightlines have physical impact parameters of $\rproj=16.9-218.4$ kpc (or $\rproj/\rvirc=0.12-0.90$) from our spiral hosts. Among the 26 sightlines, 11 were observed with both G130M and G160M gratings, while the rest have only G130M or G160M data.


For most of the archival sightlines, we download and use coadded spectra from the Hubble Advanced Spectral Products (HASP; \citealt{hasp}), which provides science-ready 1-D spectra up to program levels. Exceptions to this are sightlines \#14, \#23, and \#24 listed in Table \ref{tb:qso_info}. 
For sightline \#24, which was observed by more than one program, we use coadded spectra from the Hubble Spectroscopic Legacy Archive (HSLA; \citealt{hsla}) that combined data from all programs.  Sightlines \#14 and \#23 were observed by more than one program, and part of their data was made publicly available after the compilation of HSLA, so we coadd the spectra using the publicly available HASP spectral co-add script\footnote{https://archive.stsci.edu/missions-and-data/hst/hasp}. The coadded spectra are binned by 3 pixels to increase the signal while remaining Nyquist-sampled. The SNR of our sightlines are $\sim6-24$ (see Table \ref{tb:qso_info}). The only exception is sightline \#14, which has an SNR $=2.9$ in G130M, from which we detect a strong \HI\ Ly$\alpha$ absorber associated with the host galaxy, so we keep the \HI\ measurement in our analysis.

\begin{table}[h!]
\centering
\caption{Voigt Profile Fitting and AOD Results}
\label{tb:abs_result}
\begin{tabular}{ccccc}
\hline
Ion & $v_{\rm gal}$ & $b$ & $\log N$ & Method \\
& [$\kms$] & [$\kms$] & $[\log {\rm cm^{-2}}$] & \\ 
(1) & (2) & (3) & (4) & (5)\\
\hline
\hline 
\multicolumn{5}{c}{CSO295, NGC3432, $R_\perp=16.9$ kpc} \\
\hline
\SII\ & [-100.0, 100.0] & - & $<14.41$ & AOD\\
\CII\ &  -11.3$\pm$3.0 &   17.3$\pm$6.9 & $13.99\pm 0.12$ & Voigt\\
\CII\ &   50.4$\pm$1.8 &   32.5$\pm$4.5 & $>14.74$ & Voigt\\
\HI\ &  -11.3$\pm$3.0 &   56.4$\pm$8.7 & $>14.74$ & Voigt\\
\HI\ &   50.4$\pm$1.8 &   42.8$\pm$9.0 & $>16.09$ & Voigt\\
\OI\ &   50.4$\pm$1.8 &   59.2$\pm$18.1 & $14.90\pm 0.11$ & Voigt\\
\SiII\ &  -11.3$\pm$3.0 &   13.1$\pm$5.1 & $12.94\pm 0.09$ & Voigt\\
\SiII\ &   50.4$\pm$1.8 &   23.7$\pm$5.6 & $>13.97$ & Voigt\\
\SiIII\ &  -11.3$\pm$3.0 &   23.1$\pm$4.8 & $13.35\pm 0.11$ & Voigt\\
\SiIII\ &   50.4$\pm$1.8 &   19.4$\pm$2.9 & $>15.88$ & Voigt\\
\SiIV\ &   50.4$\pm$1.8 &   69.9$\pm$4.8 & $13.95\pm 0.02$ & Voigt\\
\hline
\multicolumn{5}{c}{PMNJ1103-2329, NGC3511, $R_\perp=86.7$ kpc} \\
\hline
\CII\ & [-100.0, 100.0] & - & $<13.38$ & AOD\\
\SII\ & [-100.0, 100.0] & - & $<14.22$ & AOD\\
\SiII\ & [-100.0, 100.0] & - & $<12.36$ & AOD\\
\SiIV\ & [-100.0, 100.0] & - & $12.91\pm 0.10$ & AOD\\
\AlII\ &  102.5$\pm$2.4 &   28.6$\pm$3.5 & $>13.29$ & Voigt\\
\CIV\ &   84.5$\pm$2.1 &   20.4$\pm$3.1 & $14.26\pm 0.05$ & Voigt\\
\HI\ &   72.2$\pm$2.1 &   58.3$\pm$3.4 & $>14.54$ & Voigt\\
\OI\ &   76.1$\pm$1.5 &   23.4$\pm$2.5 & $14.74\pm 0.04$ & Voigt\\
\SiIII\ &   81.8$\pm$2.1 &   18.0$\pm$3.3 & $13.02\pm 0.05$ & Voigt\\
\hline
\multicolumn{5}{c}{SDSSJ123657.20+265701.2, NGC4559, $R_\perp=160.3$ kpc} \\
\hline
\CII\ & [-100.0, 100.0] & - & $<13.74$ & AOD\\
\OI\ & [-100.0, 100.0] & - & $<14.34$ & AOD\\
\SII\ & [-100.0, 100.0] & - & $<14.64$ & AOD\\
\SiII\ & [-100.0, 100.0] & - & $<13.08$ & AOD\\
\SiIII\ & [-100.0, 100.0] & - & $<12.65$ & AOD\\
\SiIV\ & [-100.0, 100.0] & - & $<13.19$ & AOD\\
\HI\ &  -47.0$\pm$4.2 &   33.4$\pm$6.2 & $13.93\pm 0.08$ & Voigt\\
\hline
\hline
\end{tabular}
\begin{tablenotes}
\small \item Note: 
Col (1): Ion. 
Col (2): For results with Voigt profile fitting, $v_{\rm gal}$ indicates the fitted centroid velocity of the corresponding component in the rest frame of a target galaxy. For those with AOD measurements, $v_{\rm gal}$ shows the velocity integration range. 
Col (3): For results with Voigt profile fitting, $b$ shows the Doppler width of the corresponding component. 
Col (4): Column density measurements from either Voigt profile fitting or AOD integrations. When there is no detection, we report a $3\sigma$ upper limit value. 
Col (5): Method used to measure ion absorption. See Section \ref{sec:our_uv_sample} for details. \\
(This table is published in its entirety in the machine-readable format. A portion is shown here for guidance regarding its form and content.)
\end{tablenotes}
\end{table}

We proceed with continuum fitting following the same procedures outlined in \cite{zheng_comprehensive_2024} and briefly describe the steps as follows. We are interested in a list of absorption lines typically occurring in CGM, including Ly$\alpha$ 1215 \AA, \OI\ 1302 \AA, \SiII\ 1190/1193/1260/1526 \AA, \SiIII\ 1206 \AA, \SiIV\ 1393/1402 \AA, \CII\ 1334 \AA, \CIV\ 1548/1550 \AA, and \AlII\ 1670 \AA. For each line, we focus on a continuum spectral area within $\sim\pm1000~\kms$ from the systemic velocity of a target galaxy and normalize the absorption-line free region using the \texttt{Linetools} package \citep{prochaska16_linetools}. Given that most of our galaxies are at velocities well beyond the MW's Ly$\alpha$ absorption core, we are able to identify and fit Ly$\alpha$ absorption lines caused by \HI\ absorbers associated with the host galaxies; and most of the \HI\ absorption lines are strong and saturated. For metal ions, we only consider an absorption feature to be real when it occurs in at least two transition lines of either the same or different ions. We also analyze \FeII\ 1142/1143/1144 \AA, \SII\ 1250/1253/1259 \AA, \PII\ 1150 \AA, and \NV\ 1238/1242 \AA\ lines, but either do not find significant detection or could not reliably identify CGM absorption due to severe blending.

For \OI\ 1302 \AA, we perform an additional night-only data reduction\footnote{DayNight.ipynb, \href{https://github.com/spacetelescope}{https://github.com/spacetelescope}, N. Kerman} to remove contamination from airglow \OI\ emission from the Earth's atmosphere. We only perform night-only data reduction for sightlines where potential \OI\ absorbers associated with target galaxies fall within the airglow spectral region, and the host galaxies are found with significant \HI\ or metal detections. For each sightline, we download raw files from MAST and filter COS's TIME-TAG data in each individual exposure to only select photons taken during the night with the Sun's altitude $<0\degree$, which is below the geometric horizon from HST's point of view. We then process the filtered data using the CalCOS pipeline (version 3.6.1) and coadd all exposures (across programs if available) using HASP's spectral co-add script. In total, we process 11 (out of 26) sightlines with night-only data reduction, among which 7 sightlines yield significant SNR ($\gtrsim4$) near \OI. 

For lines with significant detections ($\geq3\sigma$), we proceed with Voigt-profile fitting using the \texttt{VoigtFit} package \citep{voigtfit}. We first isolate a spectral window of $\pm 400~\kms$ around a target galaxy's systemic velocity and visually identify velocity ranges for absorption likely associated with the galaxy halo. We then integrate the velocity ranges to estimate initial column densities and centroid velocities using the apparent optical depth method (AOD; \citealt{savage91}). These initial parameters are then used as inputs for \texttt{VoigtFit} to simultaneously fit all available lines for a given ion. During the fit, we convolve the model profiles with the COS line-spread function downloaded from MAST. In some cases, we further constrain the fit by tying the centroid velocities of different ions together. We perform the fit iteratively until the parameters column density $\log N$, centroid velocity  $v$, and Doppler width $b$ converge and the reduced $\chi^2$ value is close to 1. When there is no detection within $\pm400~\kms$, we estimate a $3\sigma$ upper-limit column density for each ion using the strongest line available by integrating the normalized line spectra over a velocity span of $\sim200~\kms$ around the systemic velocity using the AOD method. The results of our Voigt profile fitting and AOD integration are tabulated in Table \ref{tb:abs_result}.

\subsection{UV CGM Data from the Literature}
\label{sec:uv_literature}

We further collate CGM UV absorption measurements from several literature studies that include low-$z$ spiral galaxies with similar MW masses, including the COS-Halos survey \citep[$z\sim0.2$, 44 galaxies; ][]{werk_cos-halos_2013}, the CIViL* survey \citep[$z\sim0.03-0.25$, 11 galaxies;][]{garza25_civil}, and a low-$z$ CGM survey \citep{keeney17} with 10 targeted galaxies at $z\leq0.02$ and 35 serendipitous galaxies at $z\leq0.2$. For each survey, we show in Figure \ref{fig:gal_population} the corresponding stellar and halo mass ranges; note that we have recalculated the halo masses and virial radii using the same method as described in Section \ref{sec:gal_sample}. We only select star-forming spiral galaxies with $\mhalo=10^{11.5-12.7}~\msun$ from each survey to be consistent with our archival sample. 

We also include recent CGM results for three nearby MW-like spiral galaxies, including M31 \citep[54 QSO sightlines at $\rproj=11-569$ kpc; ][]{lehner_project_2020, lehner26_amiga2}, NGC 891 \citep[two sightlines at 5 and 108 kpc, respectively; ][]{qu_hstcos_2019}, and NGC 4631 \citep[one sightline at 39 kpc; ][]{richter18}. %

\begin{figure*}[t]
    \centering
    \includegraphics[width=\textwidth]{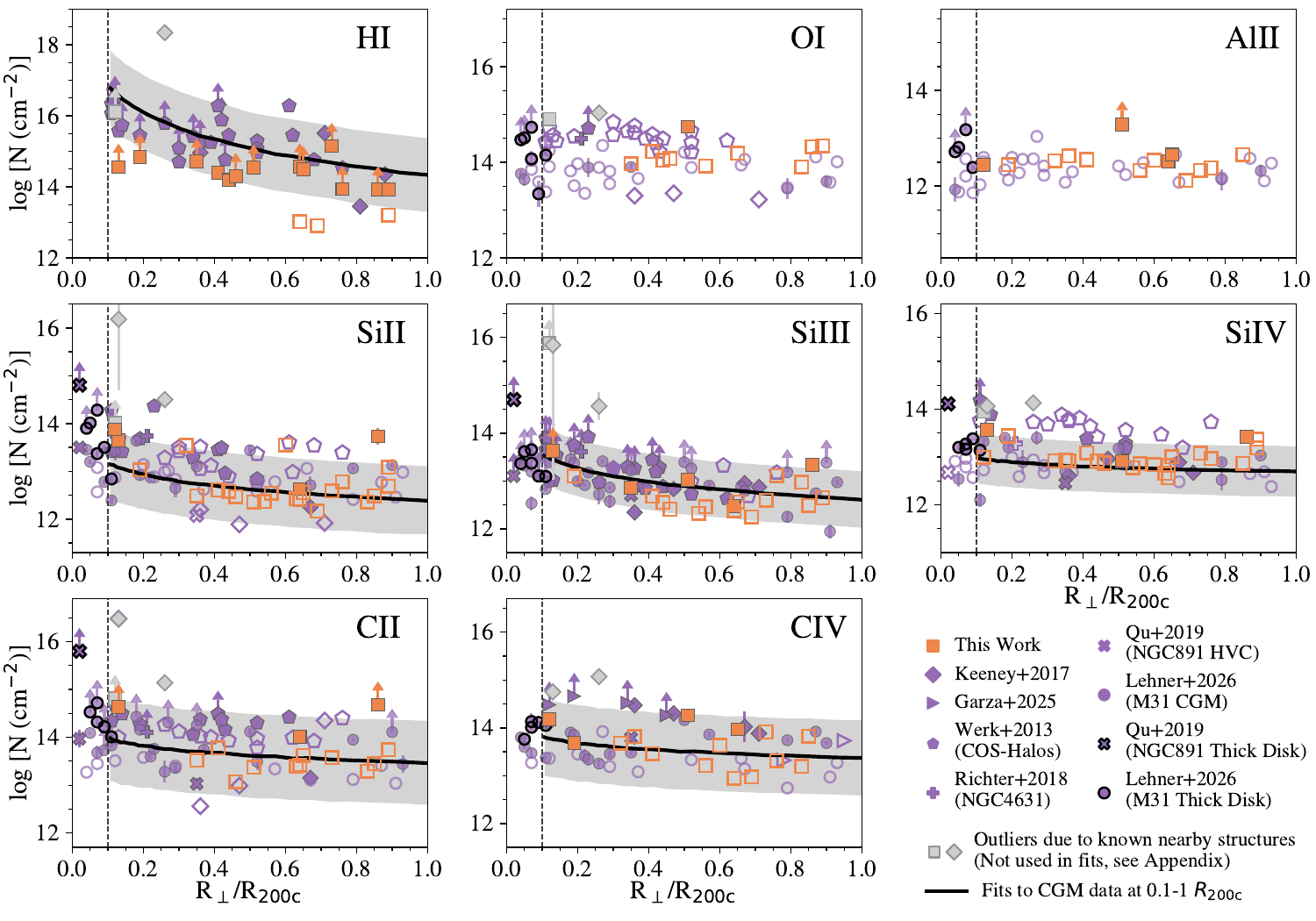}
    \caption{Total ion column density, $\log N$, as a function of impact parameter $\rproj$, normalized to each host galaxy's virial radius $\rvirc$. Detections are indicated with filled symbols and error bars, lower limits due to saturation are with filled symbols and upward arrows, and non-detection upper limits (3$\sigma$) are with open symbols. Symbols with thick black edges are associated with the respective host galaxies' thick disks and are not considered in our analysis. Gray symbols are outlier absorbers due to nearby known structures as discussed in Appendix \ref{app2:result_uni_cases}, and they are not used in our anlaysis. The thick black curves and grey shades indicate the 50th percentiles and 16th-84th confidence intervals of the \texttt{PyMC} profile fits to all the data except M31 at $0.1-1.0\rvirc$. Profile fitting results for \OI\ and \AlII\ are not constraining due to the lack of sufficient detections. 
    In general, we find that the total ion column density profiles decline as a function of $\rproj$. 
    }
    \label{fig:logN_rnorm_fit}
\end{figure*}

We only consider absorbers at $0.1\leq \rproj/\rvirc\leq1$ to avoid contamination from galaxies' extended thick disks \citep{messere26}. For M31, \cite{lehner26_amiga2} reported that there are 6 QSO sightlines within 30 kpc that have strong absorbers corotating with M31's disk, which they dubbed as the ``thick disk" component. For completeness, we show these thick disk absorbers in Figure \ref{fig:logN_rnorm_fit} with thick black boundaries, but do not use them in our analysis. Similarly, \cite{qu_hstcos_2019} found two absorbers toward a QSO sightline at $\rproj=4.7$ kpc near NGC 891's minor axis, and identified one of the absorbers to be associated with NGC 891's galactic fountain flows, while the other as a high-velocity cloud (HVC) in the galaxy's halo. We show both absorbers as crosses in Figure \ref{fig:logN_rnorm_fit}, but only use the HVC absorber in our analysis.

For each literature reference, when there is non-detection, most of these works reported a 2$\sigma$ upper limit $\log N$ value, in which case we convert the value to $3\sigma$ for consistency. The exceptions are \cite{keeney17} and \cite{richter18}, where they did not specify the significance of the upper limits, and we use their reported values without modification. It is also worth noting that \cite{werk_cos-halos_2013} and \cite{richter18} provide absorber measurements based on both AOD integration and Voigt profile fitting methods. For \cite{werk_cos-halos_2013}, we use their AOD measurements, which are the ``adopted" results throughout their analysis. For \cite{richter18}, we also use their AOD measurements that provide values for both detected absorbers and non-detected upper limits.

One of our archival sightlines, PMNJ1103-2329 (\#8 in Table \ref{tb:qso_info}), was included in the \cite{keeney17} targeted galaxy sample for the galaxy NGC 3511 (also in our galaxy sample; \#7 in Table \ref{tb:gal_info}). We keep this QSO-galaxy pair in our sample because it is one of the stronger CGM detections and represents a potential outlier case as we discuss in Appendix \ref{app2:result_uni_cases}. We confirm that our ion measurements for this sightline are consistent with values reported by \citeauthor{keeney17} within $\sim$0.1 dex. We remove this QSO-galaxy pair from the \cite{keeney17} sample in the following analysis to avoid redundancy.

Lastly, for all CGM absorbers considered in this work, if there are multiple velocity components detected in a single sightline associated with the same host galaxy (either detection or saturation lower limit), we sum up the absorbers' column densities and use the total ion column density and propagate the errors. When there are multiple upper limits for a sightline, we adopt the maximum value for that sightline as a conservative estimate. When there are both non-detection upper limits and detections along a sightline, we only use the detection values as conservative estimates for the total column density.

\section{Results}
\label{sec:result}

We first present and discuss the observed CGM ion column density profiles for local $z\sim0$ spiral galaxies compared to those of low-$z$ galaxies in Section \ref{sec:result_cgm_profile}. In Section \ref{sec:result_cgm_vs_sat} we investigate the connection between the presence of gaseous satellites and CGM detections for the host galaxies. 

\subsection{CGM Ion Profiles for Local Spiral Galaxies}
\label{sec:result_cgm_profile}

In Figure \ref{fig:logN_rnorm_fit}, we show the CGM total column density $\log N$ as a function of the normalized impact parameter $\rproj/\rvirc$ for \HI, \OI, \AlII, \SiII, \SiIII, \SiIV, \CII, and \CIV. Measurements from our archival sample are shown in orange squares, and the rest of the literature data are in purple with various symbols specified in the legend. Except for \HI\footnote{We note that M31 has \HI\ 21cm emission observations toward its QSO sightlines at various impact parameters from the Green Bank Telescope \citep{howk17}. However, none of the sightlines yield detections, resulting in a $5\sigma$ upper limit of $\log N({\rm HI})\approx17.6$. We do not show these measurements in the \HI\ panel in Figure \ref{fig:logN_rnorm_fit} because here we use UV absorption line measurements from Ly$\alpha$ 1215 \AA\ (and other higher order Lyman series lines if the absorbers are from \citealt{keeney17}). }, all ion panels span the same dynamic range of 5.2 dex along the y-axis for a straightforward comparison in the slopes and scatters in the column density profiles from ion to ion.

When available, we highlight those absorbers associated with a galaxy's thick disk with thick black edges, including one absorber from NGC 891 (cross) and absorbers from six QSO sightlines in M31 (circles). The vertical dashed line in each panel indicates the $0.1\rvirc$ boundary. It is clear that most of the thick disk components are found within $0.1\rvirc$, and they generally have higher $\log N$ values than the CGM absorbers along the same lines of sight.

\cite{lehner26_amiga2} pointed out that M31's CGM absorbers at $\rproj \lesssim0.35\rvirc$ show lower columns than those in COS-Halos at similar radii. As shown in Figure \ref{fig:logN_rnorm_fit}, we only observe such discrepancy at $\lesssim0.1\rvirc$. At $\rproj\sim0.1-0.35\rvirc$, M31's CGM absorbers do not appear to be significantly different from those spirals of similar masses in COS-Halos or other star-forming spirals in the local Universe. The discrepancy noted by \citeauthor{lehner26_amiga2} is most likely due to the higher mass or quiescent galaxies in the COS-Halos sample.

\begin{table}[b!]
\centering
\caption{Column Density Profile Fits}
\label{tb:pymc_fit}
\begin{tabular}{cccc}
\hline
Ion X & Slope $k$ & Intercept $\log N_0$ & Intrinsic Scatter $\sigma_{\rm in}$  \\
(1) & (2) & (3) & (4) \\
\hline
\HI & $-2.52^{+0.32}_{-0.33}$ & $14.34^{+0.14}_{-0.14}$ & $1.02^{+0.20}_{-0.15}$ \\
\SiII & $-0.78^{+0.15}_{-0.15}$ & $12.39^{+0.07}_{-0.07}$ & $0.72^{+0.11}_{-0.09}$ \\
\SiIII & $-0.94^{+0.12}_{-0.12}$ & $12.61^{+0.06}_{-0.06}$ & $0.59^{+0.09}_{-0.07}$ \\
\SiIV & $-0.28^{+0.13}_{-0.13}$ & $12.69^{+0.07}_{-0.07}$ & $0.52^{+0.08}_{-0.07}$ \\
\CII & $-0.52^{+0.18}_{-0.18}$ & $13.46^{+0.08}_{-0.09}$ & $0.87^{+0.16}_{-0.12}$ \\
\CIV & $-0.44^{+0.17}_{-0.18}$ & $13.37^{+0.09}_{-0.09}$ & $0.78^{+0.15}_{-0.12}$ \\
\hline
\end{tabular}
\begin{tablenotes}
\small \item Note: 
Results of \texttt{PyMC} fits to CGM absorbers at 0.1--1.0$\rvirc$. 
See Section \ref{sec:result_cgm_profile} for detail. 
\end{tablenotes}
\end{table}

We perform column density profile fits, including all CGM data points at $0.1-1.0\rvirc$, except for those outliers that are color-coded in gray; these outliers are excluded mainly due to contamination from known nearby structures, which we discuss in detail in Appendix \ref{app2:result_uni_cases}. For each ion, we fit the profile with a power-law relation, $\log N= k\log(\rproj/\rvirc)+\log N_{\rm 0}+\mathcal{N}(0, \sigma_{\rm in}^2)$, where $k$ is the profile slope, $\log N_{\rm 0}$ the intercept, and the last term $\mathcal{N}(0, \sigma_{\rm in}^2)$ a Gaussian distribution that models the intrinsic scatter $\sigma_{\rm in}$ in column densities. We follow the same fitting method using \texttt{PyMC3} \citep{pymc3} as outlined in \citeauthor{zheng_comprehensive_2024} (\citeyear{zheng_comprehensive_2024}; see their appendix C). The main difference from \citeauthor{zheng_comprehensive_2024} is that we model the intrinsic scatter as a third parameter with a uniform prior of $\sigma_{\rm in} \in\ [0.0, 5.0]$, while \citeauthor{zheng_comprehensive_2024} adopted a constant $\sigma_{\rm in}$ value based on the detected column densities. In Figure \ref{fig:logN_rnorm_fit}, we show in black curves the best-fit models (50th percentiles), and highlight in gray shades the 16th-84th percentiles (68\% confidence interval). The fitting results are tabulated in Table \ref{tb:pymc_fit}. We do not have reliable results for \OI\ and \AlII\ because most of the data are non-detections and the \texttt{PyMC} fits do not converge.

For ions with well-constrained fits, the fits produce column density profiles that decrease with impact parameter throughout the CGM. For M31, \cite{lehner26_amiga2} found that low ions, including \OI, \SiII, \SiIII, and \CII\, show stronger correlations (more negative profile slopes) between $\log N$ and $\rproj$ using the Kendall's $\tau$ test. A similar trend is found in our analysis for the silicon species, where \SiII\ and \SiIII\ show steeper profiles than the intermediately ionized \SiIV. However, we find that \CII\ and \CIV\ exhibit similar slopes within $1\sigma$.

We also note that the scatters in the $\log N$ profiles for M31's CGM are as pronounced as those in the rest of the star-forming spirals. The large spread in $\log N$ for M31 indicates that a single halo can host a diverse population of CGM clouds that lead to intrinsic scatters as large as an ensemble of CGM clouds measured from different galaxy halos.

Even though most of the \HI\ measurements are saturated, the top left panel of Figure \ref{fig:logN_rnorm_fit} shows that \HI\ has the steepest $\log N$ profile, suggesting that most of the cold gas likely resides in the inner CGM. For the three samples that have \HI\ measurements in the UV through Ly$\alpha$ or higher order Lyman series lines, including our archival sample, the COS-Halos \citep{werk_cos-halos_2013}, and \cite{keeney17}, the detection rate is nearly 100\% within $\rvirc$. In our archival sample, there are three sightlines (\#12, \#19, \#20 in Table \ref{tb:qso_info}) without \HI\ detections, which all occur at large impact parameters $\rproj>0.6\rvirc$. And for those sightlines with only G160M data (\#16, \#21, \#22, \#25), given the near-unity detection rate in \HI, it is likely that detectable \HI\ absorbers exist in their halos, should G130M data become available.  


We note that the nearly 100\% detection rate in \HI\ absorbers is not inconsistent with M31's results from \cite{howk_2017}, who found non-detections in \HI\ 21cm emission toward all of their QSO sightlines. The main reason for their non-detections is most likely that the \HI\ 21cm emission is not sensitive to absorbers with $\log N (\rm HI)\lesssim17.5$, while Ly$\alpha$ is a very strong and sensitive line with a saturated absorption profile typically at $\log N({\rm HI})\gtrsim14-14.5$. The large scatter in the \HI\ column densities presented here may also help to explain the lack of higher \HI\ column density systems in M31's halo relative to other star-forming spirals at similar masses. It is likely that these \HI\ clouds are short-lived, with a  continuous cycling between cool and hot phases \citep[e.g.,][]{porter25}.

Lastly, to compare the CGM properties at $z\sim0$ (local galaxies) and $z\sim0.2$ (low redshift), we split our archival and literature data into two subsamples and fit them with power-law column density profiles for each population separately. We do not find any significant difference in the fits for the two subsamples (not shown here). This indicates that there is either little evolution in the CGM column density profiles from $z\sim0.2$ to $z\sim0$, or the evolution is not as significant when compared to the intrinsic scatters in the $\log N$ profiles.

\subsection{The Relation between CGM and Gaseous Satellites}
\label{sec:result_cgm_vs_sat}

In this section, we investigate the connection between the presence of gaseous satellites and the properties of the host CGM. As described in Section \ref{sec:gal_sample}, we have a complete census of the relatively massive gaseous satellite population, with a typical sensitivity of $M_{\rm HI,lim} \approx 10^{7}\ M_{\odot}$. The abundance of gaseous satellites per host in our sample is overall consistent with previous studies of MW-like spirals at $z \sim 0$ (Figure \ref{fig:sat_population}).

\begin{figure}[t]
    \centering
    \includegraphics[width=\columnwidth]{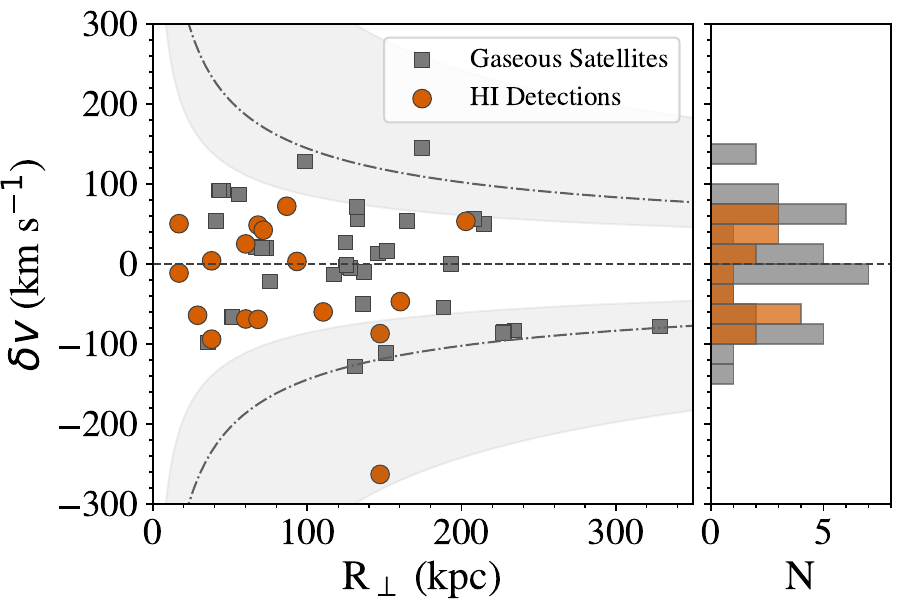}
    \caption{Velocity offset between host galaxies and \HI\ absorbers (orange dots) or gaseous satellites (gray squares) as a function of impact parameter. The dotted-dashed curves show criteria used to select gaseous satellites at a given $\rproj$ for a typical galaxy in our sample with a median halo mass of $\log (\mhalo/\msun)=11.95$ (Equation \ref{eq:vesc}), and the gray bands show the velocity range allowed by the overall halo mass range of our sample, which is $\log (\mhalo/\msun)=11.5-12.7$. All satellites and absorbers are consistent with being gravitationally bound within their respective host galaxy halos. The right panel shows the velocity distributions of the \HI\ absorbers and the gaseous satellites.}
    \label{fig:delv}
\end{figure}

In Figure \ref{fig:delv}, we show the velocity distribution of detected \HI\ absorbers around the systemic velocities of host spirals in our archival sample (red circles). Unlike in Figures \ref{fig:logN_rnorm_fit} or \ref{fig:logN_vs_Nsat} where absorber column densities are combined when there are multiple absorbers present in the same line of sight of the same halo (see Section \ref{sec:our_uv_sample}), here the velocities of individual absorbers are plotted to show the full distribution. As a comparison, we show in gray squares the velocity distribution of gaseous satellites around their host galaxies. Overall we find that both the \HI\ absorbers and the gaseous satellites have a relatively symmetrical velocity distribution around host galaxies, with the gaseous satellites showing a broader distribution out to $\sim\pm150~\kms$, while the \HI\ absorbers are mainly within $\pm100~\kms$ (see a similar trend in M31; \citealt{lehner_project_2020}). 

We note that the narrower distribution in \HI\ absorbers is not a selection effect given that we have searched for potential absorbers out to $\pm400~\kms$ from host galaxies (see Section \ref{sec:our_uv_sample}). It is possible that higher velocity absorbers exist in the halos at smaller impact parameters $\rproj\lesssim0.1-0.2\rvirc$, similar to the ionized high-velocity clouds in the Milky Way \citep{putman_gaseous_2012-2, richter18}; however, we do not have sufficient sightlines at such close proximity to host galaxies to detect those absorbers. The broadly overlapping velocity distributions motivate examining whether the host CGM column densities correlate with the gaseous satellite properties, which we present in the following subsections.

\begin{figure*}[t]
    \centering
    \includegraphics[width=\textwidth]{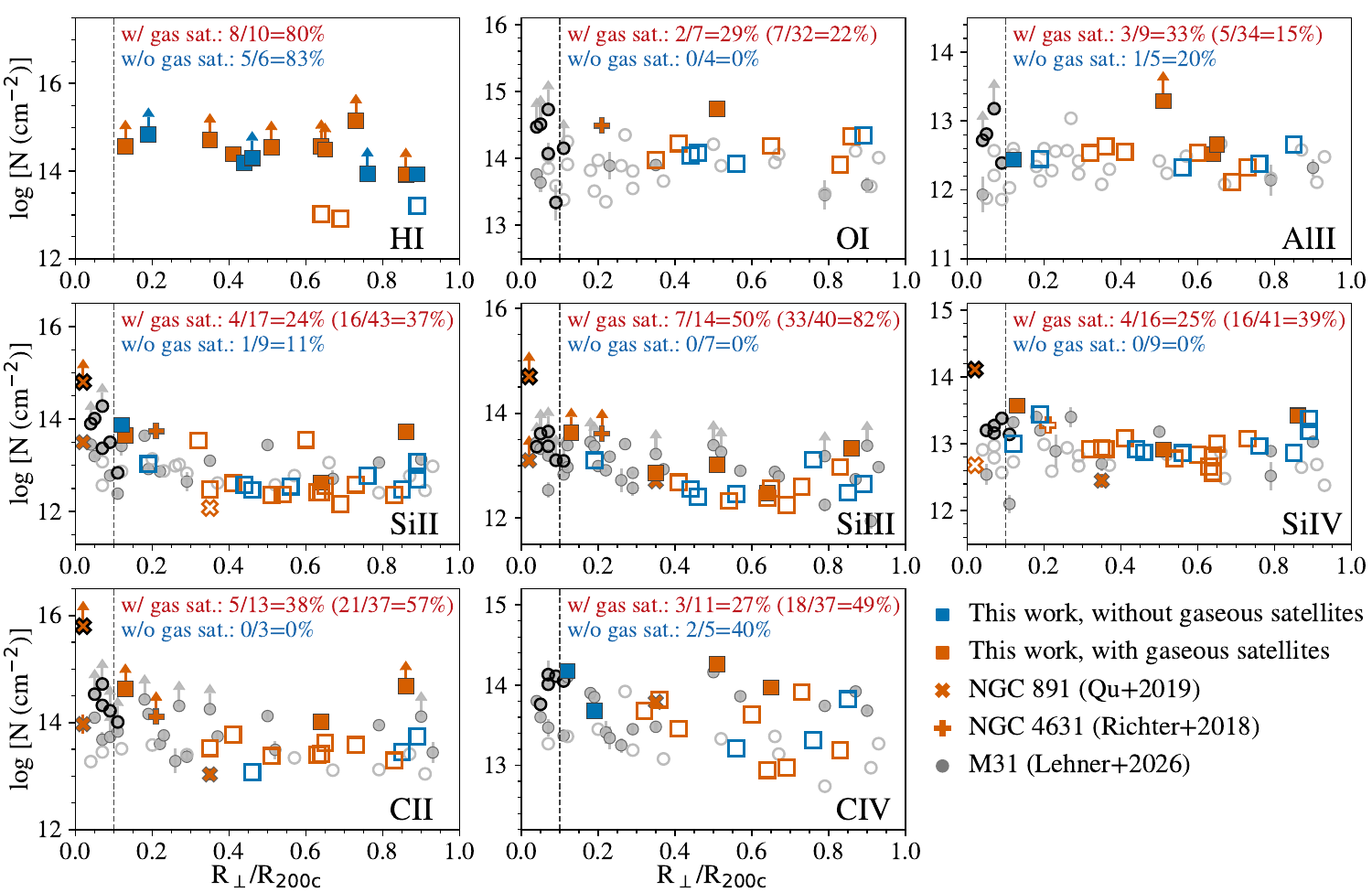}
    \caption{Ion column density \textsl{vs.} impact parameter normalized to $\rvirc$, for spiral hosts with (orange and gray) and without (blue) gaseous satellites. The three galaxies from the literature, NGC 891, NGC 4631, and M31, all have known gaseous satellite populations. For clarity, M31 data points are plotted in gray to better illustrate the detections and non-detections in other galaxy hosts. The texts in the top left corner of each panel indicate the detection rates of CGM \HI\ or metal absorbers within 0.1--1.0 $\rvirc$ for host galaxies with (orange text) and without gaseous satellites (blue text). The numbers in parentheses are the ion detection rates when M31 is considered together with the rest of the sample. We find that the spiral galaxies with and without gaseous satellites show similar detection rates of cold neutral gas in their CGM (\HI\ panel); however, when considering the ionized gas in the cool ionized phase, the CGM of spiral galaxies with gaseous satellites show considerably higher detection rates in various ions except for \CIV. }
    \label{fig:logN_w_wo_Sat}
\end{figure*}

\begin{figure}[t]
    \centering
    \includegraphics[width=\columnwidth]{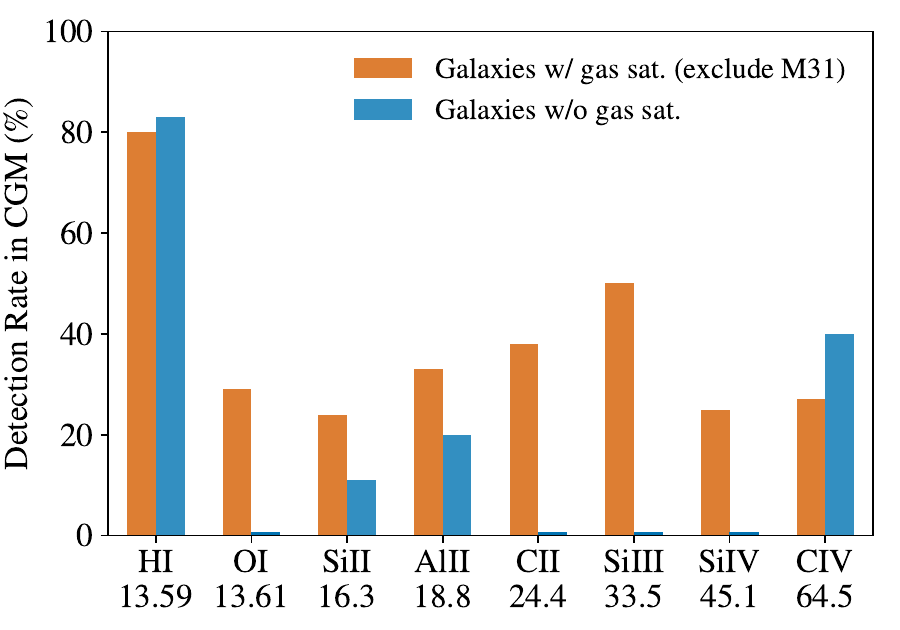}
    \caption{Ion detection rates in the CGM (0.1--1.0 $\rvirc$) for spiral galaxies with (orange) and without (blue) gaseous satellites. The numbers beneath the ion labels show the ionization potentials. The ion column density measurements and limits are shown in Figure \ref{fig:logN_w_wo_Sat}. }
    \label{fig:detection_rate}
\end{figure}

\subsubsection{CGM Detection Rates in Local Spiral Galaxies with and without Gaseous Satellites}
\label{sec:result_w_wo_sat}

We first examine whether there is a significant difference in the CGM column density properties for spiral galaxies with and without gaseous satellites. In Figure \ref{fig:logN_w_wo_Sat}, we split our galaxy sample into two subgroups, one with gaseous satellites above our \HI\ sensitivity limit ($M_{\rm HI, lim}\approx10^7~\msun$) and the other without. We also include the three local spiral galaxies, M31 (gray circles), NGC 891 (orange ``x"), and NGC 4631 (orange cross), which have known gaseous satellite populations from the literature (see more details in Section \ref{sec:result_logN_Nsat}).

In each panel, we indicate in the top left corner the detection rates (or covering fractions) of CGM absorbers for the two subgroups (without M31). Given the small number of sightlines, especially for spiral hosts without gaseous satellites, we caution that the uncertainties in the detection rates are likely high and subject to Poisson errors. We indicate in parentheses how the inclusion of M31's halo sightlines \citep{lehner26_amiga2} impacts the overall statistics for spiral hosts with gaseous satellites, but focus on analyzing the detection rates without M31 to avoid the numbers being skewed toward a single galaxy. Note that there are no \HI\ UV absorption measurements for M31. The CGM detection rates for the two subgroups are summarized in Figure \ref{fig:detection_rate}.

For \HI\ absorbers, we find that the detection rates within 0.1--1.0$\rvirc$ for galaxies with and without gaseous satellites are similar, at 80\% and 83\%, respectively. The \HI\ column density profiles between the two subgroups also appear to be similar. However, given that most of the data points are lower limits due to saturation, it is unclear whether the true \HI\ column density profiles would show discrepancies between the two subgroups.

In the \OI\ panel, we find that the spiral hosts with gaseous satellites show higher detection rates than those without (29\% \textsl{vs.} 0\%). Given that \OI\ and \HI\ trace the same cold neutral phase with similar ionization potentials (13.6 eV; Figure \ref{fig:detection_rate}), the higher detection rate in \OI, but not in \HI, for spiral hosts with gaseous satellites is likely not an indication of different ionization conditions between the two subgroups. Instead, this may suggest that spiral hosts with gaseous satellites have higher CGM column density profiles in the cold neutral phase, or that the metallicities of these host halos are higher, both of which lead to more detectable \OI\ absorbers.

For other metal ion absorbers, we find that the CGM of spiral hosts with gaseous satellites have considerably higher detection rates, up to 50\%, for metal ions including \AlII, \CII, \SiII, \SiIII, and \SiIV, which trace a cool, ionized phase of the CGM gas. In \CIV, the CGM detection rate for galaxies without gaseous satellites appears to be marginally higher, by 13\%. Figures \ref{fig:logN_w_wo_Sat} and \ref{fig:detection_rate} suggest that the presence of gaseous satellites may correlate with a higher CGM gas content in the cool, ionized phase, leading to higher detection rates. 

\begin{figure}[t]
    \centering
    \includegraphics[width=\columnwidth]{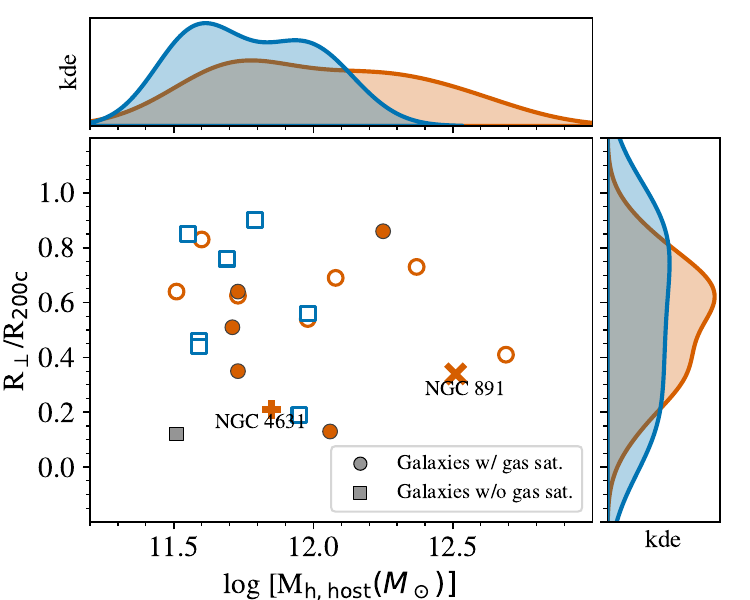}
    \caption{Impact parameters \textsl{vs.} halo masses for spiral hosts with (circles) and without (squares) gaseous satellites. Open blue/orange symbols are sightlines with no detections in \SiIII, and filled blue/orange symbols are with \SiIII\ detections. The filled gray square indicates an outlier pair (NGC3432/CSO295) with an abnormally high \SiIII\ measurement, which we do not consider in our analysis (see Appendix \ref{app2:result_uni_cases} for details). The two subgroups have similar impact parameter coverages over 0.1--1.0 $\rvirc$ (right panel), and the spiral galaxies without gaseous satellites tend to have smaller halo masses (top), by 0.3 dex on average.}
    \label{fig:mhalo_vs_r}
\end{figure}

Figure \ref{fig:mhalo_vs_r} shows the coverage of our sample sightlines in terms of the impact parameters and the halo masses for the spiral hosts with (circles) and without (squares) gaseous satellites. The right panel shows that the two subgroups have similar spatial coverage -- there is no evidence that the inner halo, which tends to have higher column densities (see Figure \ref{fig:logN_rnorm_fit}), is better sampled in one subgroup than the other.

The top panel shows that the spiral hosts without gaseous satellites tend to have lower halo masses (see also Figure \ref{fig:sat_population}). The median halo mass for the spiral hosts without gaseous satellites is $M_{\rm halo, med}=10^{11.74}~\msun$, compared to $M_{\rm halo, med}=10^{12.06}~\msun$ for the hosts with gaseous satellites. One may conclude that the lower CGM detection rates in the spiral hosts without gaseous satellites are due to their lower halo (thus CGM) masses. However, we do not find the $\sim0.3$ dex difference in $M_{\rm halo}$ to be a significant cause for the discrepancy. When only considering the lower mass galaxies with $M_{\rm halo}=10^{11.5-12.1}~\msun$, we find that the CGM detection rates in the subgroup with gaseous satellites are still consistently higher, with changes less than 9\% for all ions.

To summarize, there is a considerable discrepancy in the metal ion detection rates, but not \HI, between the CGM of spiral hosts with gaseous satellites and those without, although we note the caveat of large uncertainties due to small number of sightlines in our sample. The discrepancy is unlikely to be caused by differences in ionization conditions or halo masses. Instead, it may suggest that the CGM of spiral hosts with gaseous satellites are more metal enriched or have higher column densities. We discuss further in Section \ref{sec:discuss_sat_pop} possible causes and implications for the discrepancy in the two subgroups.

\subsubsection{CGM Column Density \textsl{v.s.} Global Properties of Gaseous Satellites}
\label{sec:result_logN_Nsat}

\begin{figure*}[t]
    \centering
    \includegraphics[width=\textwidth]{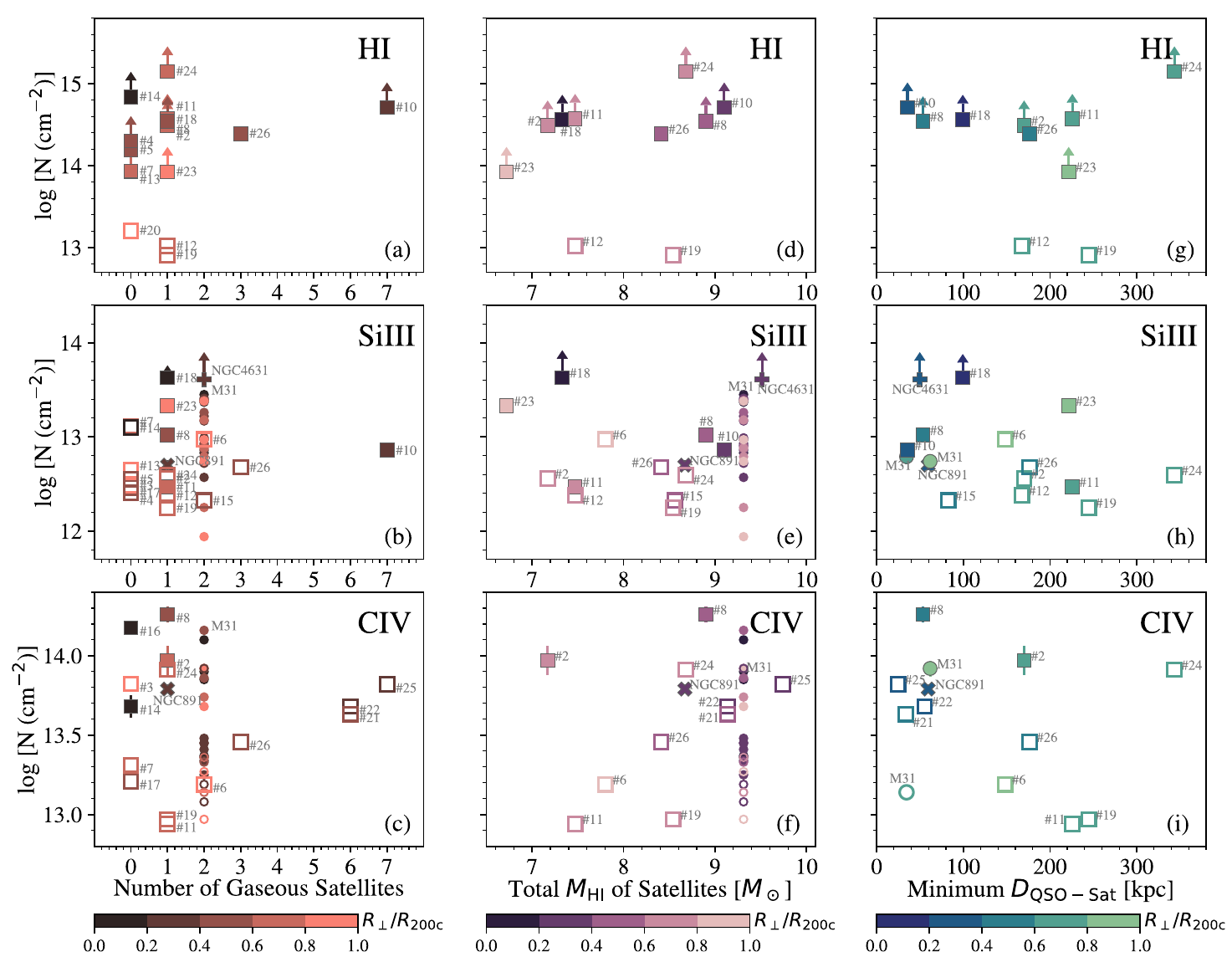}
    \caption{Ion column densities $\log N$ as functions of the numbers of gaseous (\HI\ rich) satellites within $\rvirc$ of host galaxies (left panels), total \HI\ mass contained in satellite galaxies within $\rvirc$ (middle panels), and projected distances between QSO sightlines and the nearest gaseous satellites within $\rvirc$ (right panels). The symbols are color-coded by QSO sightlines' impact parameters normalized to $\rvirc$. Galaxies with no gaseous satellites are not included in the middle and right panels. Ion detection and saturation are shown with filled symbols and non-detections are open symbols. For our sample, the numbers next to the squares correspond to the pair IDs in Table \ref{tb:qso_info}. We also include M31's CGM components (circles), NGC 891 (``x"s), and NGC 4631 (crosses). Although there is a lack of clear global correlations in the properties examined here, we discuss in Section \ref{sec:result_qso_sat_dist} that there is various evidence for higher detection rates for \HI\ and metal absorbers in the proximity of massive gaseous satellites.}
    \label{fig:logN_vs_Nsat}
\end{figure*}

In this section, we further examine, for those galaxies with gaseous satellites, how the CGM properties vary with the gaseous satellite properties, and the results are presented in Figure \ref{fig:logN_vs_Nsat}. Among all the ions available, we plot \HI\ (top row) and \SiIII\ (middle row) because they are the two strongest transitions, and \CIV\ (bottom) because it probes the warmest gas phase in our data. Although not shown, \CII, \SiII, and \SiIV\ generally exhibit similar behaviors as observed in \SiIII\ and \CIV. We also show data from M31, NGC 891, and NGC 4631.
For M31, we only include the ``CGM absorbers" at 0.1--1.0 $\rvirc$ identified in \cite{lehner26_amiga2} and consider the two satellites above our typical \HI\ mass sensitivity ($M_{\rm HI} \gtrsim 10^{7}~\msun$), M33 and IC 10.  
For NGC 891, we identify one gaseous satellite, UGC 1807 ($\log M_{\rm HI}=8.67$), using \HI\ data from the FASHI survey \citep{zhang_fast_2024}, 
and include the ``HVC" absorber at 4.7 kpc and the CGM absorber at 108.1 kpc in NGC 891's halo from \cite{qu_hstcos_2019}. For NGC 4631, we include the two gaseous satellites from \cite{karunakaran22} that are above our \HI\ sensitivity limit:   
dw1241+3251 ($\log M_{\rm HI}= 7.41$) and NGC 4656 ($\log M_{\rm HI}= 9.51$).

The left column of Figure \ref{fig:logN_vs_Nsat} shows the ion column densities as a function of the abundance of gaseous satellites within $\rvirc$. Except that galaxies with $N=0$ gaseous satellites show considerably lower detection rates in metal ion absorbers (see Figure \ref{fig:detection_rate}, Section \ref{sec:result_w_wo_sat}), there is no significant correlation between the column density and the number of gaseous satellites.

In the middle column, we examine the ion column density as a function of the total \HI\ mass contained in gaseous satellites, $M_{\rm HI, sat}^{\rm tot}$, but also find no correlation. We note that the non-correlation is not due to a lack of massive satellites in these galaxies. Out of the ten galaxies with 1--3 gaseous satellites in our sample (see Table \ref{tb:gal_info}), five galaxies (NGC 3511, NGC 5457, NGC 7814, NGC 5194, NGC 4565) have LMC-like satellites with $M_{\rm HI}>10^8~\msun$, and three of these systems have massive satellites within 50 kpc from the host galaxies (see Table \ref{tb:sat_info}). And the three galaxies from the literature, M31, NGC 891, and NGC 4631, all have LMC-like or more massive gaseous satellites in their halos. Given that the sample is already small, we do not further split it based on whether a galaxy's $M_{\rm HI, sat}^{\rm tot}$ is above or below a certain threshold mass. 

Two of the three ``outlier" host galaxies in our sample with large satellite counts (NGC 4258 has $N_{\rm HI\ sat}=6$ and NGC 3627 has $N_{\rm HI\ sat}=7$; see Figure \ref{fig:sat_population}) have no CGM detection (sightlines \#21, \#22, and \#25 in the \CIV\ panel), largely independent of the impact parameters of the sightlines from the hosts. However, these sightlines have only been observed in G160M (see Table \ref{tb:qso_info}) and we cannot rule out the possibility that there might be an \HI\ detection if G130M data were available. The third outlier host galaxy, NGC 3486 with 7 gaseous satellites, has two saturated \HI\ absorbers 
and a weak detection in \SiIII\ along sightline \#10. 
The sightline passes through the halo of NGC 3486 at an impact parameter of 60.1 kpc ($\rproj=0.35\rvirc$). We will show below that this sightline is likely associated with one of the gaseous satellites.

\subsubsection{CGM Column Density \textsl{v.s.} QSO-Satellite Distance}
\label{sec:result_qso_sat_dist}

The right column of Figure \ref{fig:logN_vs_Nsat} shows the CGM column densities as a function of the distances between QSO sightlines and the nearest gaseous satellites within $\rvirc$. Host galaxies with no gaseous satellites are not considered in this column. Below, we discuss all sightlines with ion detections within $R_{\rm \perp} \lesssim 100$ kpc of a gaseous satellite case by case. Overall, we find several cases in which sightlines at close QSO-satellite distances are likely impacted by the satellite or a known gaseous structure. For M31, for clarity, we only consider the QSO sightlines within $\sim0.5\rvirc$ of M33; there is no sightline near the other massive satellite, IC10, with the closest sightline at a projected distance of $\sim180$ kpc.

\textit{(1)} Sightline \#8 around NGC 3511 is likely impacted by its massive satellite galaxy, NGC 3513 ($\log M_{\star} \approx 9.6$, $\log M_{\rm HI} \approx 8.9$; Table \ref{tb:sat_info}). A deep \HI\ map from the MHONGOOSE survey reveals an \HI\ bridge connecting NGC 3511 and NGC 3513, confirming this system as a close, interacting pair $\sim$40 kpc apart \citep{de_blok_mhongoose_2024}. Our QSO sightline does not go through the bridge, but is closer to NGC 3513 ($\sim$53 kpc or $0.4 \rvirc$)
than to NGC 3511 ($\sim$87 kpc or $0.5 \rvirc$). We discuss this system in more detail in Appendix \ref{sec:app_ngc3511}.

\textit{(2)} Sightline \#10 around NGC 3486 is $\sim$35 kpc (or $\sim0.5R_{\rm 200c, UGC6102}$) away from its dwarf satellite UGC 6102, which is the second most massive satellite among the seven ($\log M_{\star} \approx 7.7$, $\log M_{\rm HI} \approx 8.3$; Table \ref{tb:sat_info}). The relative velocity of the satellite ($\delta v \approx 20~\kms$) is also consistent with the \HI\ and \SiIII\ detections. It is likely that sightline \#10 passes through debris associated with the satellite; however, there is no map of the extended surroundings between NGC 3486 and the satellites to confirm the existence of, e.g., stripped \HI\ clouds. We note that all satellites of NGC 3486 cluster at very close velocities around the host ($|\delta v| \leq 30~\kms$; Table \ref{tb:sat_info}), which indicates an unusually complex local environment. 

\textit{(3)} Similar to the two cases above, \cite{richter18} identifies that the sightline around NGC 4631 is associated with Spur 2, a large-scale \HI-emitting tidal feature connecting the host galaxy and one of its dwarf satellites, NGC 4656 (QSO-satellite distance is $\sim$50 kpc or $\sim0.5R_{\rm 200c,NGC4656}$; see panel h). NGC 4656 has a large \HI\ mass ($\log M_{\rm HI}=9.51$; \citealt{haynes_arecibo_2018}), 
and it is in an interacting Magellanic-like dwarf pair with NGC 4657\footnote{The NGC 4656-NGC4657 system is highly disturbed and often treated as one object (e.g., \citealt{carlsten_exploration_2022-1}).}. The high ionized-to-neutral gas ratio, low metallicity, and lack of dust in Spur 2 suggest that it is likely gas stripped from the satellite \citep{richter18}. The QSO sightline is further away ($\rproj\sim80$ kpc) from another gaseous satellite, dw1241+3251, with $M_{\rm HI}=10^{7.41}~\msun$ \citep{haynes_arecibo_2018}; given the larger distance and smaller satellite gas mass, it is unlikely that this low mass satellite would have impact on the QSO sightline detection.

\textit{(4)} For M31, we find two sightlines, 3C48.0 and RXS\_J0155.6+3115, at $R_{\rm \perp, M33}=35$ ($\sim0.2R_{\rm 200c, M33}$) and 62 ($\sim0.4R_{\rm 200c, M33}$) kpc from M33 (projected at M31's distance to be consistent with the rest of our sample) from the AMIGA survey \citep{lehner26_amiga2}, respectively. The two sightlines are at much further projected distances from M31, with $R_{\rm \perp, M31}=178$ ($\sim0.7R_{\rm 200c, M31}$) and 232 ($\sim0.9R_{\rm 200c, M31}$) kpc, respectively. When considering both the velocities and positions, \cite{kim24} show that the absorbers detected along these two sightlines have higher probabilities to be associated with M33 than M31 (see their Table 3). Given that these two sightlines are much closer to M33 than M31, and M33 shows significant surrounding structures in \HI\ emission \citep{putman_m33}, it is likely that the absorber detections are impacted by the presence of M33.

\textit{(5)} For NGC 891, \cite{qu_hstcos_2019} find two sightlines, LQAC 035+042 003 and 3C 66A, within its halo at impact parameters of 5 ($\sim0.02\rvirc$) and 108 kpc ($\sim0.34\rvirc$), respectively. The two sightlines are at $\rproj=73$ ($\sim0.8R_{\rm 200c, sat}$) and 60 ($\sim0.6R_{\rm 200c, sat}$) kpc from NGC 891's gaseous satellite, UGC 1807, which has $\log M_{\rm HI}=8.67$. Given that the sightlines are much closer to NGC 891 in projection when factoring in the size of the halo, it is likely that the detection in these two sightlines are mainly due to NGC 891's CGM and the gaseous satellite has no significant impact.


\textit{(6)} Sightline \#23 around NGC 3368 (M96) shows significant detections in \HI\ and \SiIII\ (as well as \CII, \SiII, and \SiIV) even though it is at a large impact parameter of $\rproj=0.86\rvirc$; we do not have data for \CIV\ because this sightline was only observed in G130M. The nearest dwarf satellite galaxy is too far away to explain the high column densities ($>200$ kpc). However, NGC 3368 is the primary of the M96 or Leo I group \citep{muller_leo-i_2018} and is surrounded by the Leo Ring, an intergalactic \HI\ complex more than 200 kpc in diameter \citep{stierwalt_arecibo_2009}. We find that sightline \#23 is an excellent match to AGC 205292, an \HI\ cloud within the Leo Ring characterized in the ALFALFA map of \cite{stierwalt_arecibo_2009}, both in position ($1.7 \arcmin$ or $\sim 5$ kpc) and velocity (both at $\delta v\approx 60-70~\kms$ from NGC 3368). The high column densities of the absorbers therefore reflect this dense, clumpy structure rather than the smooth, volume-filling component of the host CGM.

Lastly, we also find three cases of CGM non-detections despite close QSO-gaseous satellite distances, and they are listed as \#21, \#22, and \#25 in Table \ref{tb:qso_info}. For these three sightlines, the data are only available in G160M, so we cannot rule out the possibility that there could be detection in stronger ions such as \HI\ or \SiIII\ if G130M data exist. 



To summarize, for the spiral hosts with gaseous satellites in our sample, we do not find any global correlation between the CGM column density and either the abundance of gaseous satellites or the total \HI\ masses contained in these satellites (Section \ref{sec:result_logN_Nsat}). The CGM detection rate is only elevated when a QSO sightline is within half the virial radius of a massive gaseous satellite (LMC-like or higher mass; Section \ref{sec:result_qso_sat_dist}). It is likely that the contribution from the gaseous satellites to the CGM column density is small compared to the large intrinsic scatters in the column density profiles $\log N(\rproj)$, and that the overall trends of the $\log N(\rproj)$ profiles are mainly set by the impact parameters from the hosts (see Section \ref{sec:result_cgm_profile}). In the next section, we use an analytical model to investigate potential causes for the large intrinsic scatters in our CGM column density measurements.

\section{Analytical CGM Model - Insights and Predictions}
\label{sec:discuss_model}

To connect the observed properties of the galaxy gaseous halos to the underlying physical gas properties, we compare our data with an analytical model for cool ($T \sim 10^4$~K) CGM of MW-mass galaxies presented in \citet[][hereafter \citetalias{faerman_cool_2023}]{faerman_cool_2023}. We first briefly describe the model setup and then perform a qualitative comparison of its results and predictions with the observations, noting that the model parameters are not fitted to the individual data.

\begin{figure}
    \centering
    \includegraphics[width=\columnwidth]{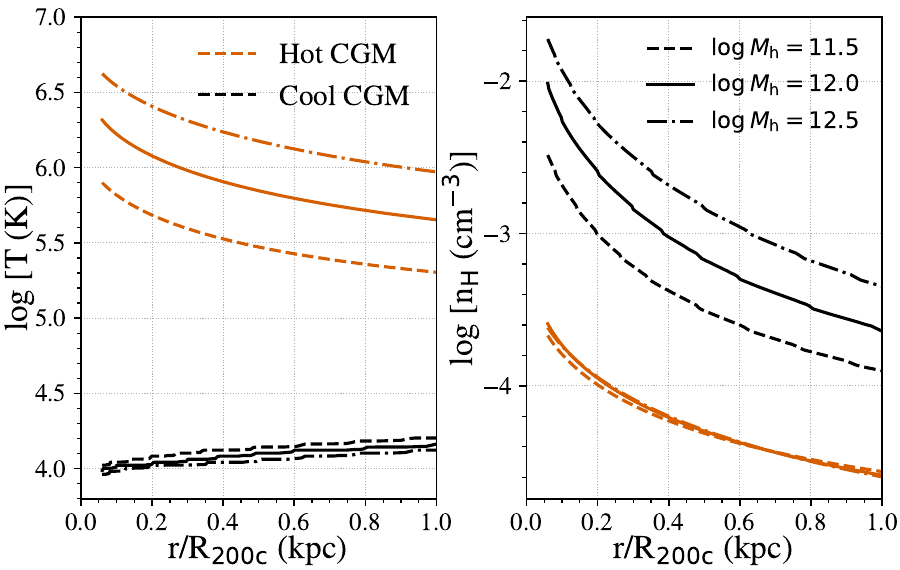}
    \caption{CGM analytic models: gas temperature (left) and hydrogen density profiles (right) for the cool (black curves) and hot (brown) phases for halo masses of $\log\mhalo$=11.5 (dashed), 12 (dotted), 12.5 (dotted-dashed, see Section \ref{sec:discuss_model} for details). The cool gas temperature is set by heating/cooling equilibrium with the MGRF, with only a weak dependence on the gas density or metallicity. The cool gas density is set by the hot gas pressure profile, and the thermal pressure contrast parameter, $\eta = P_{\rm tot}/P_{\rm th,cool} = 9$. The predicted ion column density profiles are shown in Figure \ref{fig:faerman_models}.}
    \label{fig:faerman_nr_curves}
\end{figure}

In the \citetalias{faerman_cool_2023} model, the clumpy cool phase is in total pressure equilibrium with the ambient warm/hot CGM, and the gas is in cooling/heating and photoionization equilibrium with the metagalactic radiation field (MGRF). For the ambient phase, we assume the isentropic model presented in \cite{faerman_massive_2020}. Figure \ref{fig:faerman_nr_curves} shows the CGM gas temperature (left panel) and hydrogen density (right) profiles in the models for the hot and cool phases. We construct models for three halo masses\footnote{~\cite{faerman_massive_2020} presented a model for MW-mass halos, $M_{\rm h}=10^{12}~\msun$, and here we use an extension to other halo masses described in Faerman et al. (in prep.)}: $M_{\rm h}=10^{11.5}$ (dashed), $10^{12}$ (solid), and $10^{12.5}~\msun$ (dashed-dotted), which are chosen to cover the range of halo masses in our galaxy sample ($M_{\rm h}=10^{11.5-12.7}~\msun$; see Section \ref{sec:data_method}). The temperature of the warm/hot phase (brown curves) at the virial radius is set by shock heating and varies with the halo mass as $T_{\rm hot} \propto M_{\rm h}^{2/3}$. We allow for non-thermal support in the warm/hot gas and set the CGM mass to be a constant fraction of the halo baryon budget, $M_{\rm hot} = 0.3f_{\rm bar}M_{\rm h}$ ($f_{\rm bar}=0.158$, \citealp{planck18}), similar to the fiducial model in \cite{faerman_massive_2020}. This results in a similar profile for the gas densities versus normalized radii.

\begin{figure*}[t]
    \centering
    \includegraphics[width=\textwidth]{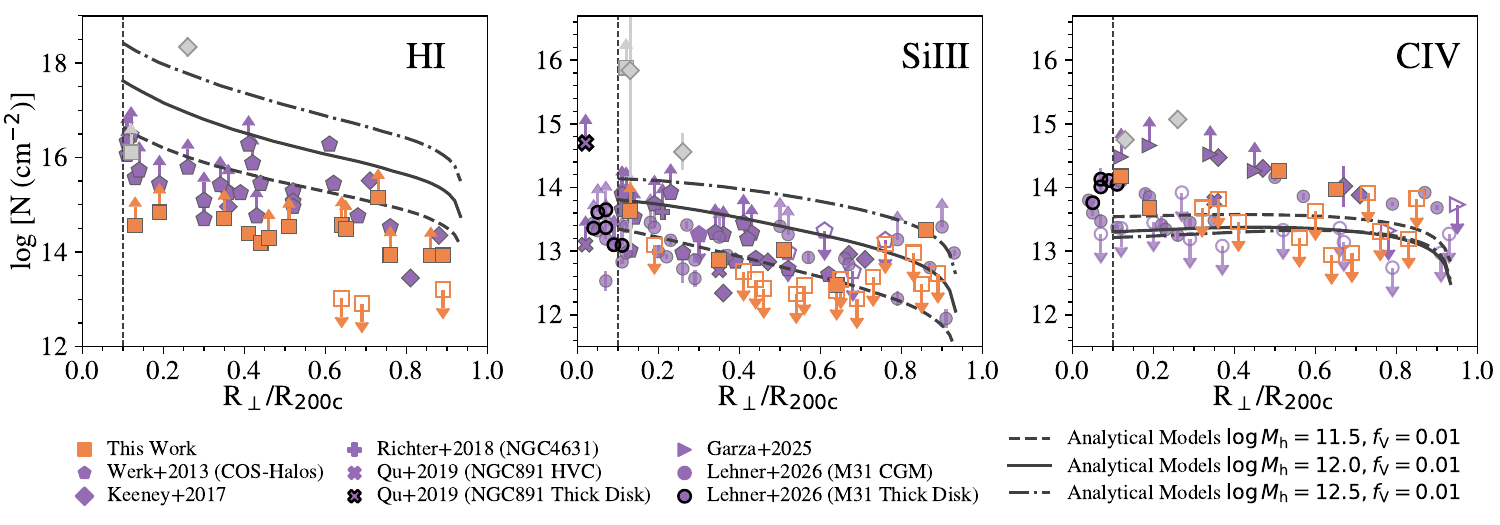}
    \caption{Total ion column densities $\log N$ as functions of the normalized impact parameter $\rproj/\rvirc$ - comparison of observations (markers, identical to Figure~\ref{fig:logN_rnorm_fit}) and analytic CGM models. We show three analytic models spanning the halo mass range of the observed galaxies, from $\log \mhalo=11.5$~(dashed) to $\log \mhalo=12.5$~(dot-dashed). The model parameters are not fitted to the observations, and the cool gas volume-filling factor is set to $f_V=1\%$, the typical value estimated in \citetalias{faerman_cool_2023}. The model predictions show varying agreement with observations, with \HI\ columns that are $1-3$~dex above the measured lower limits (left panel), to slightly higher values for \SiIII\ (middle). The models show reasonable agreement with \CIV\ observations  (right), albeit underpredicting the scatter of observed column densities. See Section \ref{sec:discuss_model} for details.}
    \label{fig:faerman_models}
\end{figure*}

For the cool phase (black curves), we calculate the gas equilibrium temperature and ionization state using Cloudy v17.00 \citep{ferland17_cloudy} assuming the \citet{khaire19_mgrf} MGRF at $z=0$. The equilibrium temperature depends weakly on the gas density and metallicity (see Figure 1 in \citealp{faerman2025}) and in our models it is in the range $T_{\rm cool} \approx (1-2) \times 10^4$~K. The model allows gas densities lower than required by thermal pressure equilibrium, either due to non-thermal support or pressure imbalance between the CGM phases. The total to thermal pressure contrast is set by the parameter $\eta = P_{\rm tot}/P_{\rm th}$ and we use a value of $\eta=9$, within the range considered in \citetalias{faerman_cool_2023}. The hot gas pressure and $\eta$ set the shape of the cool gas density profile. For the cool gas maximum extent, here we set $R_{\rm out}=\rvirc$ since our data extend to the virial radii (see Figure \ref{fig:logN_rnorm_fit}), compared to $R_{\rm out} \approx 0.6 \rvirc$ in \citetalias{faerman_cool_2023}. We adopt a volume filling fraction $f_V\equiv V_{\rm cool}/V_{\rm tot}=1$\%, similar to the value inferred by \citetalias{faerman_cool_2023} from fitting the COS-Halos ion column densities. The density profiles and $f_{\rm V}$ together determine the gas mass. For a constant volume filling fraction, the cool CGM mass in the models increases with halo mass, due to the higher gas density and larger halo volume, with $\log (M_{\rm cool}/M_\odot)=6.9 \times 10^8$, $4.8 \times 10^{9}$, and $3.0 \times 10^{10}~\msun$ for the low, middle, and high halo mass, respectively.

Figure \ref{fig:faerman_models} shows the column density profiles for \HI\ (left panel), \SiIII\ (middle), and \CIV\ (right) predicted from our analytical models, as functions of the normalized impact parameter. The columns we show are the total column densities in the CGM, a sum of the columns formed in the cool and warm/hot phases. For the low ions, the column densities in the warm/hot CGM are negligible compared to those in the cool phase for all halo masses, but as we discuss below, this is not always the case for \CIV\ and other high ions.

Overall, we find that the model column densities show trends  similar to the observational data, with the low ion column densities decreasing with $\rproj/\rvirc$ at faster rates (i.e., steeper slopes) compared to higher ions (see Section \ref{sec:result_cgm_profile}). For both \SiIII\ and \CIV, the predicted column densities have similar magnitudes to the observed columns or upper limits. For \HI, we note that most of our observational data points are lower limits from saturated Ly$\alpha$ absorption, and the true values are expected to be higher. The models predict \HI\ column densities a factor of 10-1000 higher, and these can be tested with future observations.

For \HI\ and \SiIII\ (left and middle panels), as well as \CII\ and \SiII\ (not shown here), the three models predict ion column densities spreading over $\sim1-2$ dex at a given impact parameter, similar to the level of intrinsic scatters we see in the observational data. For \SiIV, the speard in the column density profiles across the three halo mass preditions is about half of what is observed in the data. In all these ions, the slopes of the profiles do not change significantly across halo masses, but lower mass halos produce lower column densities. This suggests that the range of halo masses we examine in this work is likely to contribute to the intrinsic scatters in low ions in the cool phase.

We find a noticeable difference in \CIV\ (right panel) between the model predictions and our observational data points. The models produce a much smaller scatter in column densities compared to observations. This is set mainly by the fact that the higher mass halos ($\log M_{\rm h}=12.0$ and 12.5) have higher gas pressures and densities, which lead to lower \CIV\ ion fractions in the cool photoionized gas. These offset the higher gas densities and larger path lengths, resulting in similar column densities across halo masses in the cool phase. The low mass halo model ($\log M_{\rm h}=11.5$, dashed curve) predicts a slightly higher \CIV\ total column density profile than the two higher halo masses; this is because the temperature of the warm/hot phase at this low halo mass is close to \CIV's collisional ionization equilibrium peak, leading to a non-negligible \CIV\ column in addition to the column formed by photoionization in the cool phase.

To summarize, the \citetalias{faerman_cool_2023} model for the cool CGM, with parameters similar to their fiducial values for the COS-Halos galaxies, produces column densities that are similar to the measurements reported in this work. This is consistent with what we find in Section \ref{sec:result_cgm_profile} that the cool CGM of our galaxy sample at $z\sim0$ is not significantly different from those at $z\sim0.2$. The models show that \HI, \CII, \SiII, \SiIII, and \SiIV\ are produced mainly in the cool phase, and the halo mass range we adopt ($M_{\rm h}=10^{11.5-12.7}~\msun$) is likely to contribute to the intrinsic scatters in the ion column density profiles. For \CIV, we find that the variation due to halo mass alone cannot reproduce the observed scatters in the column density profiles, and that for low halo masses, a non-negligible fraction of the \CIV\ column densities can form in the warm/hot phase (see also discussion in \citealp{garza25b_civil}). More exploration of the model parameter space is needed to understand the amount of \CIV\ absorption and its variation in the CGM of MW-mass spirals.

\section{Discussion}
\label{sec:discussion}

We find consistency between the CGM of our $z\sim0$ spiral galaxy sample and other literature samples.  This indicates that the Milky Way's lower CGM column densities in \CIV\ \citep[e.g.,][]{bish_quastar_2021, zheng_characterizing_2020} are unlikely to be due to its current redshift. The gaseous satellites of the spiral galaxies in our sample are also consistent with the number found in previous studies. In this section, we estimate the total mass of cool CGM in the halos of spiral galaxies at $z\sim0$ in Section \ref{sec:cgm_mass} and discuss the connection between dwarf satellite populations and host CGM properties in Section \ref{sec:discuss_sat_pop}.

\subsection{Total Cool CGM Mass of $z\sim0$ Spiral Galaxies}
\label{sec:cgm_mass}

We estimate the cool CGM mass of our spiral galaxy halos by combining the column density measurements of \SiII, \SiIII, and \SiIV, following the discussion in \citealt{lehner_project_2020} and \citealt{lehner26_amiga2}. The three silicon ions cover an ionization potential ranging from 8 to 45 eV, which means they arise mostly from photoionization in the CGM and the sum of these ions traces the bulk of the cool CGM metal mass.

We estimate the mass for a typical L$_*$ galaxy in our sample, with a median halo mass of $M_{\rm h}=10^{11.95}~\msun$, which corresponds to $\rvirc=204$ kpc (see Table \ref{tb:gal_info}). By integrating the best-fit \SiII, \SiIII, and \SiIV\ column density profiles (see Table \ref{tb:pymc_fit}) from 0.1 to 1.0 $\rvirc$, we find the total silicon mass in the cool phase to be: 
\begin{equation}
    \begin{split}
        M_{\rm Si}^{\rm cool} & = \int_{0.1\rvirc}^{\rvirc} 2\pi \rproj N_{\rm SiII+SiIII+SiIV}(\rproj) m_{\rm Si}d\rproj\\
        & = 4.8^{+14.6}_{-3.6}\times10^5~M_\odot, 
    \end{split}
    \label{eq:m_si}
\end{equation}
where $m_{\rm Si}$ is the mass of a silicon atom and $R_{\perp}$ is the projected distance (i.e., impact parameter) of a sightline. The positive and negative errors are calculated directly from the 16th-84th percentiles from the \texttt{PyMC3} posterior samples. We find that the uncertainties in the slope $k$ and intercept $\log N_0$ contribute only $\sim5-22\%$ ($\pm0.8\times10^5~\msun$) of the errors, and the rest is mainly due to the large intrinsic scatter that we account for during the profile fits.

The total metal mass in the cool phase is: 
\begin{equation}
    \begin{split}
        M^{\rm cool}_{\rm Z} & = \frac{M^{\rm cool}_{\rm Z}}{M^{\rm cool}_{\rm H}} \frac{M^{\rm cool}_{\rm H}}{M^{\rm cool}_{\rm Si}} M^{\rm cool}_{\rm Si}  = Z_{\rm m} \mu_{\rm Si} M^{\rm cool}_{\rm Si} \\
        & =7.5^{+22.6}_{-5.6}\times10^6 ~\msun, 
    \end{split}
    \label{eq:m_metal}
\end{equation}
where $Z_{\rm m}=Z'~Z_{\rm m, \odot}$ is the mass fraction of metals out of all gas, for the solar value we adopt $Z_{\rm m, \odot} \approx 0.014$ (\citealp{asplund09}), and $Z'$ is the metallicity relative to the Solar value. For silicon, $\mu_{\rm Si} \equiv M_{\rm H}^{\rm cool}/M_{\rm Si}^{\rm cool}=m_{\rm H}n_{\rm H}/(m_{\rm Si}n_{\rm Si})=1/[28Z' (\frac{n_{\rm Si}}{n_{\rm H}})_\odot] = 1/(0.0009Z')$, where $(n_{\rm Si}/n_{\rm H})_\odot=10^{7.51-12}$ is the solar abundance. For $M_{\rm Z}^{\rm cool}$, the metallicity factor in $Z_{\rm m}$ and $\mu_{\rm Si}$ cancel out and does not require an assumption on a particular $Z'$ value. 

We further estimate the total CGM gas mass in the cool phase as: 
\begin{equation}
    \begin{split}
        M^{\rm cool} & = 1.4 \frac{M^{\rm cool}_{\rm H}}{M^{\rm cool}_{\rm Si}} M^{\rm cool}_{\rm Si}\\
        & =2.5^{+7.5}_{-1.9}\times10^9~\msun~(\frac{Z'}{0.3~Z_\odot})^{-1}, 
    \end{split}
    \label{eq:mcool}
\end{equation}
where we assume $Z'=0.3Z_\odot$, which is a typical metallicity value for the CGM of low-redshift galaxies \citep{prochaska_cos-halos_2017}. We note that the cool gas mass for the median halo mass analytic model described in Section~\ref{sec:discuss_model}, $M^{\rm cool} = 4.8 \times 10^9~\msun$, is within the range estimated here, providing a consistency check for these two results.

When comparing to literature values, \cite{lehner26_amiga2} find that the total metal mass of M31's cool CGM integrated over $\rproj=5-230$ kpc is $M_{\rm Z}^{\rm cool}=(1.36\pm0.82)\times10^{7}~\msun$ (or $M^{\rm cool}\approx4.5\times10^9~\msun$ assuming a CGM metallicity of 0.3$Z_\odot$; see their Table 8), which is about a factor of two higher than our estimate in Equations \ref{eq:m_metal} and \ref{eq:mcool}. There are likely to be a few factors that cause the discrepancy. First, we use a different integration range of $0.1-1.0\rvirc$, which corresponds to $\rproj=20.4-204$ kpc for a typical L$_*$ galaxy in our sample. And, as \cite{lehner26_amiga2} pointed out, what fitting method is used, how censored data are treated, and whether the intrinsic scatter is taken into account, all affect the final mass values and error estimates. Our fitting method is most similar to their second approach, which is an MCMC power-law fit that considers the likelihood of the censored data (both upper and lower limits) using cumulative distribution functions (see their Appendix C).

We note that both our mass estimates and \citeauthor{lehner26_amiga2}'s are significantly lower than the values for the COS-Halos sample at $z\approx 0.2$ inferred by \cite{werk_cos-halos_2014} ($M^{\rm cool}\sim6.5\times10^{10}~\msun$) and \cite{prochaska_cos-halos_2017} ($M^{\rm cool}\sim9\times10^{10}~\msun$). \cite{lehner26_amiga2} pointed out that for the COS-Halos sample, 44\% of the sightlines at $\lesssim0.35\rvirc$ are associated with massive halos ($M_{\rm 200}>10^{12.5}~\msun$)\footnote{We do not see these higher column density systems in our $\log N$ radial profiles in Figure \ref{fig:logN_rnorm_fit} because we only include star-forming galaxies from the COS-Halos sample with $M_{\rm h}=10^{11.5-12.7}~\msun$.}, which skew the profile fits to steeper slopes. Indeed, \citetalias{faerman_cool_2023} show that for star-forming COS-Halos galaxies, a fiducial model with  $M^{\rm cool} = 3 \times 10^{9}~\msun$ (spherical, enclosed within $R<160~{\rm kpc} = 0.6R_{\rm vir}$) can reproduce a large fraction of the measured column densities, and models with $M^{\rm cool}=(1-10) \times 10^{9}~\msun$ reproduce most of the measurements except for the highest column densities.  

\subsection{Gaseous Satellites and Host Halo Gas}
\label{sec:discuss_sat_pop}

We split our galaxy sample into two subgroups based on whether or not a galaxy host has gaseous satellites within its halo, and find considerably higher detection rates in CGM metal ion absorbers, including \OI, \AlII, \CII, \SiII, \SiIII, and \SiIV, in the subgroup of spiral hosts with gaseous satellites, by up to 50\% (Section \ref{sec:result_w_wo_sat}). The detection rates for \HI\ absorbers are similar between the two subgroups. The positive correlation between the presence of gaseous satellites and the higher CGM metal detection rates may suggest that the gaseous satellites are ``metal polluters" in their host halos, through processes such as stellar feedback or stripping of the satellites' metal-enriched CGM gas and potentially partial stripping of their outer ISM gas.

This is consistent with previous observational and simulation studies, which find that dwarf galaxies are efficient at ejecting metals through stellar feedback due to their shallow graviational potentials \citep[e.g.][]{kirby13, mcquinn15, christensen16, christensen18, andersson23}. HST/COS observations of nearby dwarf galaxies in relatively isolated environments show that their cool CGM only contain $\sim10\%$ of the total metals every produced, with the rest likely in hotter phases or escaped beyond the CGM \citep{zheng_comprehensive_2024}. \cite{piacitelli2025} also find a dearth of metals in the CGM of isolated dwarf galaxies in the Marvel-ous Dwarfs and Marvelous Massive Dwarfs hydrodynamic simulations. For gaseous satellites in our sample, it is likely that the CGM of these satellite galaxies have been (partially) stripped, and the metals these galaxies produced have been mixed into the ambient host halo media.

When considering the spiral hosts with gaseous satellites, we find no global correlation between the host CGM detections and the number of gaseous satellites (Figure \ref{fig:logN_vs_Nsat}, left column). This may partly reflect that our \HI-selected satellite census is only sensitive to relatively massive, gas-bearing satellite galaxies with $M_{\rm HI}\gtrsim10^{7}~\msun$ (see Table \ref{tb:gal_info}), and the CGM profiles of our $z\sim 0$ spiral hosts are broadly comparable to those of $z\sim0.2$ spiral galaxies (see Section \ref{sec:result_cgm_profile}). Such a host CGM is not expected to efficiently remove the ISM of satellites more massive than $M_{\star} \approx 10^{7}\ \msun$ or $M_{\rm h} \approx 10^{10}\ \msun$ through ram pressure \citep{zhu_too_2026}\footnote{This mass threshold is comparable to our \HI\ sensitivity limit assuming an \HI-to-stellar mass ratio of $\sim 1$ for satellite dwarf galaxies \citep{putman_gas_2021,zhu_baryonic_2025}.}, which have deeper gravitational potentials \citep[see also][]{salem_ram_2015-1}. A typical host CGM, along with $\leq3$ surviving gas-bearing satellites above $M_{\rm HI} \approx 10^{7}\ \msun$ (Figure \ref{fig:gal_population}), is therefore consistent with the expected efficiency of environmental stripping within MW-mass spiral halos.

Since stripping of a massive gaseous satellite's ISM is unlikely to be effective (as discussed above), the observed satellite abundance across hosts may largely reflect the typical scatter at the bright end of the satellite luminosity function, which is influenced by variations in the halo assembly history (e.g., \citealt{smercina_relating_2022,mao_saga_2024,mutlu-pakdil_faint_2024}), rather than host-to-host difference in stripping efficiency.

For the 9 hosts in our sample with targeted optical data (Section \ref{sec:gal_sample}), the satellite quenched fractions based on the ELVES survey \citep{carlsten_exploration_2022-1} are $\sim 50-70 \%$, largely consistent with the quenched fractions in the broader ELVES sample. The quenched fraction per host shows no dependence on CGM detections. Deep optical data for our full sample --- capturing both gaseous and gas-poor (quenched) satellites and constraining their stellar masses and distances --- would help improve the existing statistics and allow further assessment of how the host CGM affects the quenched fraction and gas depletion of its satellites.

We also find no correlation between the ion column densities and the total \HI\ mass contained in gaseous satellites (Figure \ref{fig:logN_vs_Nsat}, middle column). Since our \HI\ census is complete at the massive end, the total \HI\ mass captures most of the neutral gas reservoir bound to the satellite population except the stripped debris, which does not predict the CGM column densities within the same halo. Using the TNG50 simulation, \cite{damle25} find a positive correlation between the total mass of the satellite population and the host's cold CGM mass\footnote{We note that the cold CGM mass is not directly comparable with our sightline column density observable in the cool phase.}, but also find that the overall satellite contribution to the cold CGM mass budget is likely small, and much of the cold CGM lies far from the satellites.

The lack of global trends suggests that the satellite contribution to the measured CGM column densities is likely small compared to the intrinsic scatters in the column density profiles. Nevertheless, among the five cases where a QSO sightline lies within $\sim 100$ kpc of a gaseous satellite (Figure \ref{fig:logN_vs_Nsat}, right column), four show ion detections within $\sim0.5R_{\rm 200c,sat}$ of massive gaseous satellites (LMC-like or higher masses; see Section \ref{sec:result_qso_sat_dist}). Strong absorption is also found near the Leo Ring, a large-scale \HI\ structure around NGC 3368. These cases suggest that satellites and related gas structures can contribute locally to the observed CGM column densities.

Ram pressure stripping provides one possible source of satellite-associated gas in the CGM. Simulations show that stripped satellite ISM can cool and survive as clouds in host halos \citep{tonnesen_its_2021,roy24,roy_survive_2026,augustin_foggie_2025} and remain clustered around the satellites \citep{ramesh23}. For sufficiently massive satellites, nearby absorption may also arise from the satellite's own perturbed CGM, as observed around the LMC \citep{fox14, krishnarao_observations_2022,mishra_truncated_2024}. M33 may represent a similar case in the M31 halo \citep{kim24}.

Tidal interactions provide another possible source. Tidal gas streams, usually produced by close gravitational encounters between a massive satellite and either the host or another satellite, are uncommon \citep{sancisi_cold_2008,besla_frequency_2018}. But when they do occur, such streams can contribute $M_{\rm HI} \sim 10^{8}-10^{9}\ \msun$ of neutral gas to the host halo \citep[e.g.,][]{putman_gaseous_2012-2}. As shown by the case studies in Section \ref{sec:result_qso_sat_dist}, sightlines passing near tidal gas structures can show high column densities. Because tidal streams are dense and clumpy, their limited spatial extent also helps explain why only a small number of sightlines are affected. Deep, resolved \HI\ emission maps are particularly useful for identifying these structures.

\section{Summary}
\label{sec:summary}

We investigate the connection between the CGM column density profiles of MW-like spiral hosts at $z\sim0$ and the presence of gaseous satellites within their halos. Our sample includes 26 galaxy-QSO archival pairs probing 21 unique spiral hosts, observed with HST/COS G130M and/or G160M gratings, at impact parameters ranging from 0.1 to 0.9$\rvirc$ (Section \ref{sec:data_method}). We focus on MW-mass spirals with stellar masses of $M_*=10^{9.5-10.9}~\msun$ (halo masses of $M_{\rm h}=10^{11.5-12.7}~\msun$) within 15 Mpc (Figure \ref{fig:gal_population}), and compare the sample with six literature studies of spiral galaxies of similar masses. Our findings are summarized as follows. 

\begin{itemize}
    \item For \HI, \SiII, \SiIII, \SiIV, \CII, and \CIV, the ion column densities decrease with the impact parameters, consistent with previous results in the literature (Figure \ref{fig:logN_rnorm_fit}; Section \ref{sec:result_cgm_profile}). \HI\ exhibits the steepest profile, suggesting that most of the cold neutral gas is likely concentrated near the inner CGM; and \HI\ absorbers are ubiquitously detected throughout the halos, with an almost 100\% detection rate at all radii. For metal ions, \SiII\ and \SiIII\ appear to have steeper profiles than \SiIV, while \CII\ and \CIV\ profiles show similar slopes within $1\sigma$ (Table \ref{tb:pymc_fit}). Most of the \OI\ and \AlII\ data are non-detections, so the radial trends in their column density profiles are unclear.
    
    \item All ion column density profiles show large intrinsic scatters of $\sim1-2$ dex (Figure \ref{fig:logN_rnorm_fit}; Section \ref{sec:result_cgm_profile}). When splitting the data into a local sample ($z\sim0$) and a low-$z$ sample ($z\sim0.2$), we do not find any significant difference between the two sub-samples. This suggests that there is either little evolution in the CGM column density profile from $z\sim0.2$ to $z\sim0$, or the evolution is not as significant compared to the intrinsic scatters in the ion column density profiles. 
    
    \item We compare the observed ion column density profiles with predictions from an analytical model for the cool, photoionized CGM of MW-mass halos from \cite{faerman_cool_2023}. We find that the model can reproduce the column densities in \HI, \CII, \SiII, \SiIII, and \SiIV, with parameters similar to the FW23 fiducial model, and that the over-a-decade halo mass range ($\mhalo=10^{11.5-12.7}~\msun$) we adopt in this work is likely to contribute to the intrinsic scatters in the column densities of these ions (Figure \ref{fig:faerman_models}; Section \ref{sec:discuss_model}). However, the models underpredict both the magnitude and the scatter in \CIV\ column densities, suggesting the need for a more detailed exploration of the parameter space in the analytical model.
    
    \item We estimate the total metal mass in the CGM of our spiral galaxy sample between 0.1 and 1.0 $\rvirc$ to be $7.5^{+22.6}_{-5.6}\times10^6~\msun$, which corresponds to a total cool gas mass of $2.5^{+7.5}_{-1.9}\times10^9~\msun$, assuming a CGM metallicity of $Z'=0.3Z_\odot$, consistent with the estimate for M31's CGM \citep{lehner26_amiga2} when factoring in different fitting methods and radial integration ranges (Section \ref{sec:cgm_mass}). It is also consistent with the results from \cite{faerman_cool_2023}, which shows that an analytical model with $M^{\rm cool}=3\times10^9~\msun$ can successfully reproduce the CGM column densities of star-forming galaxies in the COS-Halos survey. 
    
    \item For the 21 local galaxies in our archival sample, we conduct a complete census of gaseous (\HI-bearing) satellites within their halos using wide-field \HI\ surveys with an \HI\ sensitivity of $M_{\rm HI, lim}\gtrsim10^{7}~\msun$ (Figure \ref{fig:sat_population}; Section \ref{sec:gal_sample}). Most galaxy hosts have $\leq3$ gaseous satellites, consistent with previous satellite surveys. Three galaxies, NGC 3486, NGC 4258, and NGC 3627, are with unusually high gaseous satellite counts ($N_{\rm sat}=6-7$), with the latter two being galaxy group centrals with more complex environments.
    
    \item We investigate the relation between the CGM properties and the presence of gaseous satellites. When considering the line of sight velocities, we find that both the \HI\ absorbers and the gaseous satellites have a nearly symmetrical velocity distribution around the host galaxies (Figure \ref{fig:delv}; Section \ref{sec:result_cgm_vs_sat}). The gaseous satellites show velocity offsets from host galaxies within $\sim\pm150~\kms$, while \HI\ absorbers exhibit a slightly narrower distribution within $\sim\pm100~\kms$. 

    \item We divide the spiral hosts at $z\sim0$, including our archival sample and 3 galaxies from the literature (NGC 891, NGC 4631, and M31), into two subgroups based on whether or not a galaxy host has gaseous satellites above our \HI\ mass sensitivity. We find that the detection rate of \HI\ absorbers is similar between these two subgroups (Figures \ref{fig:logN_w_wo_Sat} and \ref{fig:detection_rate}). The metal ion absorbers, including \OI, \AlII, \CII, \SiII, \SiIII, and \SiIV, on the other hand, show considerably higher detection rates in the CGM of spiral hosts with gaseous satellites, by up to 50\%. The discrepancy is unlikely to be caused by differences in ionization conditions or halo masses (Section \ref{sec:result_w_wo_sat}). Instead, it is likely that the CGM of the spiral hosts with gaseous satellites are more metal enriched or have more ionized gas in the cool phase, with contributions from the gaseous satellites through ejected metals from stellar feedback or stripped CGM gas.
    
    \item When considering the spiral hosts with gaseous satellites, we find no significant correlation between the CGM column densities and the abundance of gaseous satellites or the total \HI\ masses contained in the satellites (Figure \ref{fig:logN_vs_Nsat}). The contribution from the gaseous satellites to the global CGM column densities is likely small compared to the intrinsic scatters, and the column density profiles are largely influenced by sightlines' proximity to host galaxies. We find that the detection rates of \HI\ and metal ion absorbers are elevated when a sightline is within half the virial radius of a massive gaseous satellite (LMC-like or higher mass), likely due to stripped debris.
    
\end{itemize}

Lastly, we caution that the findings related to gaseous satellites in this work have significant uncertainties due to the currently available sample size, and more systems are needed to better quantify the relation between gaseous satellites and the CGM. In particular, high-sensitivity, wide-coverage \HI\ or future UV/optical emission maps would help understand the origins of ion absorbers near gaseous satellites and their interplay/interaction with the host CGM.

\begin{acknowledgments}
This research has used NASA's Astrophysics Data System and the analyses were based on observations obtained with the NASA/ESA Hubble Space Telescope, retrieved from the Mikulski Archive for Space Telescopes (MAST) at the Space Telescope Science Institute (STScI). Support for HST-AR-17562 was provided by NASA through a grant from STScI. STScI is operated by the Association of Universities for Research in Astronomy, Inc., under NASA contract NAS5-26555. This research has used the HSLA and HASP database, developed and maintained at STScI, Baltimore, USA.

\end{acknowledgments}





%
\facilities{Hubble Space Telescope/Cosmic Origins Spectrograph; Mikulski Archive for Space Telescopes; The Parkes telescope is part of the Australia Telescope which is funded by the Commonwealth of Australia for operation as a National Facility managed by CSIRO.}

\software{Astropy \citep{astropy:2013, astropy:2018, astropy:2022}, Numpy \citep{numpy}, Matplotlib \citep{matplotlib}, Linetools \citep{prochaska16_linetools}, VoigtFit \citep{voigtfit}, PyMC3 \citep{pymc3}}


\appendix
\restartappendixnumbering

\section{Catalog of \HI-bearing Satellite Galaxies}\label{app3:gaseous_sats}
Table \ref{tb:sat_info} shows the properties of the gaseous satellites around our spiral host galaxies from our \HI\ census, as described in Section \ref{sec:gal_sample}. 

\begin{deluxetable*}{rcccccccccc}
\tabletypesize{\footnotesize}
\tablecaption{Gaseous Satellite Information \label{tb:sat_info}}  
\tablehead{
    \colhead{GID} &
    \colhead{Galaxy} &
    \colhead{Satellite} &
    \colhead{Survey} & 
    \colhead{RA$_{\rm Sat}$} &
    \colhead{Dec$_{\rm Sat}$} &
    \colhead{$v_{\rm helio, Sat}$} &
    \colhead{$d_{\rm proj}$} &
    \colhead{$\Delta v_{\odot}$} &
    \colhead{$W_{\rm 50}$} & 
    \colhead{$\log M_{\rm HI}$} \\
& & & & [deg] & [deg] & [$\kms$] & [kpc] & [$\kms$] & [$\kms$]  & [$\log\msun$]\\
\colhead{(1)} & \colhead{(2)} & \colhead{(3)} & \colhead{(4)} & \colhead{(5)}  & \colhead{(6)} & \colhead{(7)} & \colhead{(8)} & \colhead{(9)} & \colhead{(10)} & \colhead{(11)}  
 }
\startdata 
 2 &    NGC4517 &           AGC 225760 &    ALFALFA & 187.274 & 0.101 & 1198 & 132.2 & 71 & 25 & 7.17\\
\hline
 5 &    NGC5364 &             UGC 8838 &    ALFALFA & 208.907 & 4.990 & 1170 & 35.7 & -98 & 77 & 7.46\\
   &            &           AGC 238709 &    ALFALFA & 208.473 & 4.884 & 1281 & 145.6 & 13 & 26 & 7.53\\
\hline
 7 &    NGC3511 &             NGC 3513 &  MHONGOOSE & 165.942 & -23.242 & 1191 & 43.0 & 92 & 80 & 8.90\\
\hline
 8 &    NGC4274 &             UGC 7440 &    ALFALFA & 185.609 & 29.204 & 904 & 117.1 & -13 & 155 & 7.47\\
\hline
 9 &    NGC3486 &           AGC 722731 &    ALFALFA & 164.933 & 28.607 & 698 & 66.8 & 21 & 12 & 6.57\\
   &            &             UGC 6102 &    ALFALFA & 165.449 & 28.689 & 697 & 70.7 & 20 & 70 & 8.31\\
   &            &           AGC 208569 &    ALFALFA & 164.695 & 29.220 & 697 & 73.1 & 20 & 81 & 7.11\\
   &            &           AGC 212945 &    ALFALFA & 165.165 & 29.708 & 673 & 124.8 & -4 & 44 & 7.52\\
   &            &             UGC 6126 &    ALFALFA & 165.937 & 28.888 & 704 & 124.8 & 27 & 184 & 8.98\\
   &            &           AGC 219369 &    ALFALFA & 165.963 & 28.686 & 667 & 137.0 & -10 & 22 & 7.30\\
   &            &           AGC 215232 &    ALFALFA & 165.310 & 29.849 & 693 & 151.5 & 16 & 23 & 7.23\\
\hline
12 &    NGC5457 &  J140535.09+542711.3 &       FAST & 211.396 & 54.453 & 297 & 41.0 & 54 & 64 & 8.56\\
   &            &  J135406.44+534652.8 &       FAST & 208.527 & 53.781 & 296 & 164.6 & 54 & 22 & 6.52\\
\hline
14 &    NGC3351 &           AGC 200532 &    ALFALFA & 160.503 & 12.332 & 772 & 128.0 & -5 & 36 & 7.33\\
\hline
15 &    NGC7814 &            UGC 12918 &    ALFALFA & 0.515 & 16.593 & 1050 & 125.4 & -1 & 102 & 8.54\\
\hline
17 &    NGC4258 &  J121811.01+465458.5 &       FAST & 184.546 & 46.916 & 396 & 51.5 & -66 & 62 & 7.58\\
   &            &  J122038.64+461720.6 &       FAST & 185.161 & 46.289 & 518 & 132.6 & 56 & 210 & 8.74\\
   &            &  J121014.66+473406.8 &       FAST & 182.561 & 47.569 & 408 & 188.2 & -54 & 23 & 6.78\\
   &            &  J122114.41+454902.6 &       FAST & 185.310 & 45.817 & 462 & 193.2 & 0 & 28 & 8.13\\
   &            &  J121728.71+453652.5 &       FAST & 184.370 & 45.615 & 512 & 214.7 & 50 & 117 & 8.78\\
   &            &  J122953.89+473152.4 &       FAST & 187.475 & 47.531 & 379 & 234.3 & -83 & 41 & 7.65\\
\hline
18 &    NGC3368 &           AGC 202024 &    ALFALFA & 161.251 & 11.913 & 871 & 75.7 & -22 & 24 & 6.72\\
\hline
19 &    NGC5194 &  J133403.43+475440.6 &       FAST & 203.514 & 47.911 & 353 & 150.8 & -112 & 125 & 8.68\\
\hline
20 &    NGC3627 &             UGC 6328 &    ALFALFA & 169.733 & 13.095 & 803 & 55.5 & 87 & 493 & 8.40\\
   &            &             UGC 6350 &    ALFALFA & 170.061 & 13.589 & 844 & 98.6 & 128 & 458 & 9.69\\
   &            &           AGC 210220 &    ALFALFA & 169.253 & 13.098 & 588 & 130.7 & -128 & 25 & 7.10\\
   &            &           AGC 213440 &    ALFALFA & 170.908 & 12.896 & 666 & 136.0 & -50 & 22 & 6.72\\
   &            &           AGC 202257 &    ALFALFA & 169.810 & 11.953 & 861 & 174.6 & 145 & 51 & 7.83\\
   &            &           AGC 202256 &    ALFALFA & 168.689 & 12.650 & 630 & 226.7 & -86 & 42 & 7.15\\
   &            &             UGC 6272 &    ALFALFA & 168.650 & 12.814 & 631 & 227.7 & -85 & 254 & 8.38\\
\hline
21 &    NGC4565 &             UGC 7758 &    ALFALFA & 188.896 & 25.850 & 1353 & 45.6 & 92 & 130 & 8.32\\
   &            &           AGC 732226 &    ALFALFA & 190.200 & 25.945 & 1317 & 208.0 & 56 & 35 & 7.33\\
   &            &           AGC 732179 &    ALFALFA & 188.840 & 27.554 & 1183 & 328.7 & -78 & 93 & 7.45\\
\hline
\enddata
\tablenotetext{}{
Note: 
Col (1): Galaxy ID, same as in Table \ref{tb:gal_info}. 
Col (2): Galaxy name. 
Col (3): Gaseous satellite name. 
Col (4): Survey name.
Col (5): RA of the satellite. 
Col (6): Dec of the satellite. 
Col (7): Heliocentric velocity of the satellite. 
Col (8): Satellite-host on-sky projected distance (adopting host galaxy distance in Table \ref{tb:gal_info}). 
Col (9): Satellite-host heliocentric velocity offset. 
Col (10): \HI\ velocity linewidth (full width at half maximum).  
Col (11): \HI\ mass scaled to the host galaxy's distance as adopted in Table \ref{tb:gal_info}.  
} 
\end{deluxetable*}

\section{Notes on Some Unique Galaxy-QSO Pair Cases}
\label{app2:result_uni_cases}

In this section, we discuss a few unique galaxy-QSO pair cases that have higher than usual column densities, as shown in Figure \ref{fig:logN_rnorm_fit}. 

\subsection{NGC2611/PG 0832+251 from \citeauthor{keeney17}}

In the \HI\ panel, we find a sightline from \cite{keeney17} with $\log N=18.34$ at $\rproj/\rvirc=0.26$ (or $\rproj=53$ kpc). This is from QSO PG 0832+251 in the halo of galaxy NGC 2611 from \citeauthor{keeney17}'s targeted galaxy sample. Along the sightline, the authors identified two \HI\ velocity components at $\delta v=5~\kms$ ($\log N_{\rm HI}=18.34$) and $\delta v=111~\kms$ ($\log N_{\rm HI}=14.96$) from NGC 2611, respectively. The data point in Figure \ref{fig:logN_rnorm_fit} shows the combined value of the two. As noted by \citeauthor{keeney17}, NGC 2611 is a luminous starburst galaxy. While PG 0832+251 was initially used to target the halo of NGC 2611, subsequent follow-up spectroscopy shows that NGC 2611 resides in a small group of galaxies, and the QSO sightline is, in fact, closer to another galaxy in the group, albeit fainter than NGC 2611 (although the authors did not specify which galaxy that is). It is most likely that the high \HI\ column density is caused by the complex group galaxy environment that the QSO sightline is probing, rather than solely the CGM of NGC 2611. We do not include this sightline in our \texttt{PyMC} profile fits. Note that this sightline also leads to the only detection of \OI\ from the \citeauthor{keeney17} sample in the \OI\ panel.

\subsection{M108/SBS1108+560 from \citeauthor{keeney17}}

We note that in the \SiII\ and \SiIII\ panels, there is an outlier data point with large error bars from the \cite{keeney17} sample at $\rproj/\rvir=0.13$ ($\rproj=22$ kpc) in the halo of M108 along QSO sightline SBS1108+560. For \SiII, the data point is a combination of two velocity components at $\delta v=-40$ and $20~\kms$ from the host galaxy's systemic velocities with $\log N_{\rm SiII}=16.18\pm1.5$ and $14.12\pm0.28$, respectively. For \SiIII, the data point is a combination of three components at $\delta v=-53$, 22 and $98~\kms$ with $\log N_{\rm SiIII}=13.97\pm1.88$, $15.83\pm2.87$, and $13.40\pm0.74$, respectively. It is unclear why this sightline has such high \SiII\ and \SiIII\ column densities besides the fact that it is at a small impact parameter. Given the large uncertainties and apparently higher than usual column densities, we exclude the data point in our \texttt{PyMC} fits for \SiII\ and \SiIII, which would otherwise result in slightly steeper column density profiles.

\subsection{NGC3511/PMNJ1103-2329 in Our Archival Sample}
\label{sec:app_ngc3511}
In the \AlII\ panel, the saturated value at $\rproj/\rvir\sim0.5$ is caused by one of our archival sightlines, PMNJ1103-2329 (\# 10 in Table \ref{tb:qso_info}) at  $\rproj=86.7$ kpc from galaxy NGC 3511 (\#7 in Table \ref{tb:gal_info}). As noted in Section \ref{sec:uv_literature}, this QSO-Galaxy pair was also included in the \cite{keeney17} sample. Along the same sightline, we also detect significant \HI, \OI, \SiIII\ and \CIV\ absorptions, which we fit with Voigt profiles and show the results in Table \ref{tb:abs_result}. We also find marginal detection in \NV\ with an integrated AOD $\log N_{\rm NV}=13.8$. Our measurements are generally consistent with \citeauthor{keeney17}'s within 0.1 dex. However, \citeauthor{keeney17} did not report detection in \AlII; it is unclear whether they did not have any detection in \AlII\ or this ion was not included in their survey (since there is no data for \AlII\ in any of their sightlines).

\subsection{NGC 3432/CSO295 in Our Archival Sample}
Another interesting pair in our archival sample is NGC 3432/CSO295 (\#1 in the galaxy Table \ref{tb:gal_info} and \#2 in the QSO Table \ref{tb:qso_info}), which shows strong detections in \HI, \OI, \SiII, \SiIII, \SiIV, and \CII\ (see Table \ref{tb:abs_result}) at an impact parameter of $\rproj/\rvirc=0.12$. Unfortunately, there is no G160M archival data for this sightline, so it is unclear whether other ions such as \CIV\ and \AlII\ might also be present. We are able to identify two velocity components in \HI, \CII, \SiII\ and \SiIII\ associated with this galaxy, while the \OI\ and \SiIV\ data are too noisy to yield reliable two component fits. The galaxy is nearly edge-on with an inclination angle of 85$\degree$, and the QSO sightline lies close to the minor axis of the galaxy at $\rproj=16.9$ kpc. In this case, the high column densities in multiple ions are most likely due to the proximity of the QSO sightline to the host galaxy, and that it may be probing some outflowing gas along the galaxy's minor axis. This sightline is not included in our profile fits.


\bibliography{main}{}
\bibliographystyle{aasjournalv7}


\end{CJK*}
\end{document}